\documentclass[11pt,epsf]{article}
\DeclareMathAlphabet{\scr}{U}{rsfs}{m}{n}
\usepackage{latexsym}
\usepackage{epsfig}
\usepackage[mathscr]{eucal}
\usepackage{amsfonts}
\usepackage{amscd}
\usepackage{cite}
\usepackage{array}   
\usepackage{amssymb}
\usepackage{colordvi}
\usepackage[centertags]{amsmath}
\usepackage{enumerate}
\usepackage{graphicx}
\usepackage{booktabs}
\usepackage{theorem}
\usepackage[footnotesize]{caption}
\usepackage{soul}

\newcommand{\newc}{\newcommand}
\newc{\be}{\begin{equation}}
\newc{\ee}{\end{equation}}
\newc{\bea}{\begin{eqnarray}}
\newc{\eea}{\end{eqnarray}}
\newc{\ol}{\overline}
\newc{\wt}{\widetilde}
\newc{\bs}{\boldsymbol}
\newc{\m}{\mathcal}
\newc{\la}{\langle}
\newc{\ra}{\rangle}

\newcommand{\lsim}{\raisebox{-0.13cm}{~\shortstack{$<$ \\[-0.07cm] $\sim$}}~} 
\newcommand{\gsim}{\raisebox{-0.13cm}{~\shortstack{$>$ \\[-0.07cm] $\sim$}}~}

\newcommand{\non}{\nonumber} 
\newcommand{\beq}{\begin{eqnarray}} 
\newcommand{\eeq}{\end{eqnarray}} 
\newcommand{\s}{\smallskip}

\newcommand{\sn}{\newline \vspace*{-3.5mm}}

\newcommand{\bc}{\begin{center}}
\newcommand{\ec}{\end{center}}

\newcommand{\ba}{\begin{array}}
\newcommand{\ea}{\end{array}}

\begin{document}

\title{\hfill ~\\[-30mm]
\vspace*{1cm}
\vspace{13mm}   \textbf{Natural NMSSM Higgs Bosons\\[4mm]}}

\date{}
\author{
S.~F.~King$^{1\,}$\footnote{E-mail: \texttt{king@soton.ac.uk}},
M. M\"{u}hlleitner$^{2\,}$\footnote{E-mail: \texttt{maggie@particle.uni-karlsruhe.de}},
R.~Nevzorov$^{3\,}$\footnote{E-mail: \texttt{nevzorov@itep.ru}},
K. Walz$^{2\,}$\footnote{E-mail: \texttt{kwalz@particle.uni-karlsruhe.de}}
\\[9mm]
{\small\it
$^1$School of Physics and Astronomy,
University of Southampton,}\\
{\small\it Southampton, SO17 1BJ, U.K.}\\[3mm]
{\small\it
$^2$Institute for Theoretical Physics, Karlsruhe Institute of Technology,} \\
{\small\it 76128 Karlsruhe, Germany.}\\[3mm]
{\small\it
$^3$Institute for Theoretical and Experimental Physics, Moscow, 117218, Russia.}\\
}

\maketitle

\begin{abstract}
\noindent
We study the phenomenology of Higgs bosons close to 126~GeV
within the scale invariant unconstrained next-to-minimal supersymmetric Standard
Model (NMSSM), focusing on the regions of parameter space favoured by
low fine-tuning considerations, namely stop masses of order 400~GeV to
1~TeV and an effective $\mu$ parameter between 100--200~GeV, with large
(but perturbative) $\lambda$ and low $\tan \beta =$2--4. 
We perform scans over the above parameter space, focusing on the
observable Higgs cross sections into $\gamma \gamma$, $WW$,  $ZZ$,
$bb$, $\tau \tau$ final states, and study the correlations between these observables.
We show that the $\gamma \gamma$ signal strength may be enhanced up to
a factor of about two not only due to the effect of singlet-doublet
mixing, which occurs more often when the 126~GeV Higgs boson is the
next-to-lightest CP-even one, but also due to light stops (and to a lesser
extent light chargino and charged Higgs loops). There may be also
smaller enhancements in the Higgs decay channels into $WW$,  $ZZ$,
correlated with the $\gamma \gamma$ enhancement. However there is no
such correlation observed involving the Higgs decay channels into $bb$, $\tau \tau$.
The requirement of having perturbative couplings up to the GUT scale
favours the interpretation of the 126~GeV Higgs boson as being
the second lightest NMSSM CP-even state, 
which can decay into pairs of lighter neutralinos, CP-even or CP-odd Higgs
bosons, leading to characteristic signatures of the NMSSM.
In a non-negligible part of the parameter range the increase in the
$\gamma\gamma$ rate is due to the superposition of rates from nearly
degenerate Higgs bosons. Resolving these Higgs bosons would rule out the Standard Model,
and provide evidence for the NMSSM.
\end{abstract}
\thispagestyle{empty}
\vfill
\newpage
\setcounter{page}{1}

\section{Introduction}
The recent discovery of a new particle
with a mass around $\sim125$~GeV~\cite{:2012gk,:2012gu}
is consistent with the Standard Model (SM) Higgs boson.
In particular the observed decays and signal strengths into $\gamma \gamma$, $WW$,  $ZZ$ favour the interpretation that the particle is a neutral boson with spin-0. However more data is needed to assess its nature,
and if careful studies of the signal strengths in different channels reveal discrepancies from the
predictions of the SM then this would provide
a window into new physics Beyond the Standard Model (BSM) \cite{higgsfits}.
Supersymmetric (SUSY) models are a leading candidate for BSM physics and 
generically predict one or more light Higgs bosons whose properties may
differ in detail from that of the SM Higgs boson.
For example, if the cross section of Higgs production and decay into $\gamma \gamma$
were observed to be significantly higher than the SM Higgs prediction, then this could be due to 
the effects of SUSY particles in the loops~\cite{ggsquarkcontr,Djouadi:1998az,carenastau,chargedloops,Higgs,squarkandsuppr,bbsupprandloop1,bbsupprandloop2,bbsupprandloop3,bbsupprandloop4}
or suppressed couplings to $b$-quarks leading to smaller total 
widths~\cite{squarkandsuppr,bbsupprandloop1,bbsupprandloop2,bbsupprandloop3,bbsupprandloop4,originalsingletdoublet,125singletdoublet,Ellwanger:2011aa,singletdoubletfurther,bbsuppressandfinetune,King:2012is,bbsuppress,comparative,relicdensity1,relicdensity2,relicdensity}.\sn

In the Minimal Supersymmetric Standard Model (MSSM) there exists an
upper limit on the lightest Higgs boson mass of about 130--135~GeV,
depending on the values of the parameters in the stop sector (see
{\it e.g.}~\cite{Djouadi:2005gj} and references therein). The MSSM can be
consistent with a 126~GeV SM-like Higgs boson in the decoupling
limit. In this limit the lightest Higgs boson mass is given by 
\begin{equation}
m_h^2 \, \approx  \, M_Z^2 \cos^2 2 \beta + \Delta m_h^2, 
\label{eq:hmassMSSM}
\end{equation}
with the correction $ \Delta m_h^2$ being dominated by loops of heavy top quarks and
top squarks. The ratio of the vacuum expectation values (VEVs) of the
two Higgs doublets introduced in the MSSM Higgs sector is denoted by
$\tan \beta$. In order to raise the Higgs boson mass to 126~GeV, we hence
need at large values of $\tan \beta$ a loop contribution of $\Delta
m_h \approx 85$~GeV which is nearly as large as the tree-level
mass value. This leads to some degree of fine-tuning \cite{SusyFineTuning}.\sn

It has been known for some time that the fine-tuning of the
MSSM could be ameliorated in the scale invariant Next-to-Minimal Supersymmetric
Standard Model (NMSSM)~\cite{NMSSMtuning1,125NMSSMtuning1}. With a
126~GeV Higgs boson, due to the  fine-tuning of the MSSM, the NMSSM
has emerged as a more natural alternative.
In the NMSSM~\cite{genNMSSM1,genNMSSM2,other-non-minimal1,Nevzorov:2004ge}
(for reviews see~\cite{Ellwanger:2009dp,Maniatis:2009re}) one singlet
superfield $S$ is added to the spectrum of the MSSM.
The supersymmetric Higgs mass parameter $\mu$ is then generated
dynamically through the coupling term $\lambda S H_d H_u$. The upper mass
bound of the lightest Higgs boson in the NMSSM becomes,
\begin{equation}
m_h^2 \, \approx  \,  M_Z^2 \cos^2 2 \beta + \frac{\lambda^2 v^2}{2} \sin^2 2 \beta +  \Delta m_h^2,
\label{eq:hmassNMSSM}
\end{equation}
where $v=246$~GeV.  Contrary to the MSSM, for $\lambda v > M_Z$, the tree-level
contributions to $m_h$ are maximized for moderate values of $\tan \beta$. 
For example, setting $\lambda =0.6$ and $\tan\beta=2$, these tree-level
contributions raise the Higgs boson mass to about 100~GeV requiring
$\Delta m_h \sim 75$~GeV in order to match the 126~GeV Higgs mass
value. The difference to the correction needed in the MSSM (numerically about
10~GeV) is significant as $\Delta m_h$ raises logarithmically with the
stop masses and receives an important contribution from the stop mixing.\sn

In the NMSSM, depending on $\tan\beta$, $\lambda \sim 0.7$ is the largest value in order not to
spoil the validity of perturbation theory up to the GUT scale. 
The presence of additional extra matter, however, allows larger values of
$\lambda $ to be achieved~\cite{Masip:1998jc}. 
For example, adding three families of $5+\overline 5$ extra matter at a mass scale of 1~TeV 
increases the largest value to $\lambda \sim 0.8$ for the same
parameters as before.  The above discussion shows that there is an argument from fine-tuning
for extending the NMSSM to include extra matter. Such an NMSSM+ model
with extra matter has recently been discussed in \cite{Hall:2012mx}. \sn

In this paper we study the phenomenology of Higgs bosons in the mass range
124-127 GeV within the scale invariant NMSSM. To distinguish our
study from the many NMSSM studies in the literature of a near 126~GeV
Higgs boson, we shall focus exclusively on the regions of parameter space
favoured by low fine-tuning considerations, namely stop masses of
order 400~GeV to 1~TeV and an effective $\mu$ parameter between
100-200 GeV, with large (but perturbative) $\lambda$ and low $\tan
\beta =$2--4. We shall allow for the possibility of extending the NMSSM
to include extra matter as in the NMSSM+ \cite{Hall:2012mx}, so that
$\lambda$ can be increased up to 0.8 at low energy scales, while remaining 
perturbative up to the GUT scale.  We perform scans over the above
parameter space, focusing on the observable Higgs cross sections into 
$\gamma \gamma$, $WW$,  $ZZ$, $bb$, $\tau \tau$ final states, and study the
correlations between these observables. We show that the $\gamma
\gamma$ signal strength may be enhanced up to a factor of two due to
an enhancement of the production and/or the decay mechanism. While small
stop mixing for light stops enhances the dominant production process
through gluon fusion and suppresses the loop-mediated decay into
photons, the latter can also be enhanced to 
some extent through light chargino and charged Higgs boson
loops. Furthermore, the suppression of the dominant decay into bottom
quarks due to singlet-doublet mixing entails a suppressed total
width and hence enhances the branching ratio into photons. Since also the
branching ratios into $WW$ and $ZZ$ are affected by such a suppression there
is a strong correlation between these channels and the $\gamma\gamma$ enhancement.
However there is no such correlation observed involving the Higgs
decay channels into $bb$, $\tau \tau$ which may be suppressed in the
enhanced $\gamma \gamma$ region. The results include the possible
presence of a second Higgs boson in 
the region of 126~GeV. The superposition of the rates of two nearly
degenerate Higgs bosons also increases the event rate in the photon
final state. Provided that such a signal can be disentangled from a
singly produced Higgs boson in future, this will be a further strong
test of the NMSSM.
Our scan reveals that with the chosen small stop mass values it is
difficult to get a lightest CP-even Higgs boson with mass around
126~GeV if in addition perturbativity constraints are imposed. It
can only be achieved for large mixing values and inclusion of extra
matter at $\sim 1$~TeV, while this is not necessary if the second
lightest CP-even Higgs boson is demanded to have the same mass as the
recently discovered new resonance. Interestingly, in this case the
mass spectrum can be such that the heavier CP-even Higgs boson can
decay into a pair of lightest neutralinos, CP-odd or CP-even Higgs
bosons, which leads to distinctive final state signatures to be tested
at the LHC.
\sn

The work in this paper complements and goes beyond the other studies
of a Higgs boson near 126~GeV in the NMSSM~\cite{125NMSSMtuning1,125singletdoublet,bbsupprandloop2,bbsupprandloop3,bbsupprandloop4,originalsingletdoublet,King:2012is,constrainednmssm,comparative,relicdensity1,relicdensity2,Graf:2012hh,Gunion:2012gc,relicdensity,Kyae:2012rv,Bae:2012am,Gogoladze:2012jp,SchmidtHoberg:2012ip}.
For example the original observation that the di-photon channel may be
enhanced due to strong singlet-doublet mixing due to the reduction of
the $bb$ partial width with a second lighter CP-even Higgs
boson was made in \cite{originalsingletdoublet}. This was followed by
our proposal \cite{King:2012is} of a set of benchmark points in which  
we studied, in addition to $\gamma \gamma$ and $bb$,
also the channels $WW$ and $ZZ$ both for the case where the second
CP-even Higgs boson is lighter or heavier than the 126~GeV one, for the case
of light top squarks and gluinos.  
The channels were also studied in the framework of various versions of
the constrained NMSSM with relatively heavy stops
\cite{constrainednmssm}, and a comparative study between the MSSM and
NMSSM has been performed in \cite{comparative}. In
\cite{relicdensity1} the effect of astrophysical and Dark Matter (DM) 
constraints on the lightest supersymmetric particle (LSP) in the NMSSM
with a 126~GeV Higgs boson was taken into account with the main focus on the
LSP being a singlino-like neutralino. Similar constraints were also
applied to the constrained NMSSM with a dominantly Higgsino-like LSP \cite{relicdensity2}. 
The case of the Higgs boson mass spectrum in the complex NMSSM  was
considered in \cite{Graf:2012hh} leading to significant effects on
Higgs phenomenology. In \cite{Gunion:2012gc} scenarios were 
investigated where the two lightest NMSSM Higgs bosons are closely
spaced near 126~GeV, leading to very enhanced decay rates, and in
\cite{relicdensity} scenarios with Higgs bosons both consistent with
the LEP 98~GeV excess and the 126~GeV Higgs boson of the LHC search
were studied,
while \cite{bbsupprandloop3} discussed the case where the NMSSM Higgs sector
could both explain the 126~GeV discovery and the small excess
observed by CMS at 136~GeV.
In \cite{Kyae:2012rv} a more complicated non-NMSSM model with extra
singlets was proposed, while in  \cite{Bae:2012am} another alternative
to the NMSSM involving singlet mass terms was studied. The case of
fine-tuning in the NMSSM was analysed in \cite{NMSSMtuning1}.
Finally, the effects of combining the NMSSM with an inverse see-saw mechanism
were considered in \cite{Gogoladze:2012jp}. \sn

The layout of the remainder of this paper is as follows. In
section~\ref{sec:nmssm} we briefly introduce the Higgs sector of the
scale invariant NMSSM. In section~\ref{sec:scan} we present in detail the
parameter values which we choose for the scan in the NMSSM parameter
space. This is followed, in section~\ref{sec:proddec}, by the
discussion of the SUSY particle effects 
in the loop mediated processes of the dominant NMSSM Higgs production
through gluon fusion on the one hand and the decay into a photon pair
on the other hand. Section~\ref{sec:numerics} contains the numerical
analysis with the presentation and discussion of $\lambda$-$\kappa$ and
mass distributions, of total widths, branching ratios and reduced
rates with their correlations. A comparison with the present LHC Higgs
search results is presented. Section~\ref{sec:concl} summarises and
concludes the paper.
 

\section{The NMSSM \label{sec:nmssm}}  
We restrict ourselves to the NMSSM with a scale invariant superpotential. We do not take
into account other possible extensions as the minimal
non-minimal supersymmetric SM (MNSSM), new minimally-extended supersymmetric SM
or nearly-minimal supersymmetric SM (nMSSM), neither extensions with additional $U(1)'$ gauge
symmetries \cite{other-non-minimal2}, nor the case of explicit CP
violation \cite{cpvnmssm,Graf:2012hh,cpvnmssmmass}. \sn 

Including only the third generation fermions, the NMSSM superpotential
in terms of (hatted) superfields is given by
\beq
{\cal W} = \lambda \widehat{S} \widehat{H}_u \widehat{H}_d +
\frac{\kappa}{3} \, \widehat{S}^3 + h_t
\widehat{Q}_3\widehat{H}_u\widehat{t}_R^c - h_b \widehat{Q}_3
\widehat{H}_d\widehat{b}_R^c  - h_\tau \widehat{L}_3 \widehat{H}_d
\widehat{\tau}_R^c \; .
\label{supot2}
\eeq
The first term replaces the $\mu$-term $\mu
\widehat H_u  \widehat H_d$ of the MSSM superpotential, while the
second  one, cubic in the singlet superfield, is introduced to break
the Peccei-Quinn symmetry \cite{pqsymm} 
in order to avoid the appearance of a massless axion. The last three terms
represent the Yukawa interactions. The scalar mass parameters for the Higgs
and sfermion scalar fields which contribute to the soft SUSY breaking
Lagrangian read in terms of the fields corresponding to the complex scalar
components of the superfields, 
\beq
\label{Lmass}
 -{\cal L}_{\mathrm{mass}} &=& 
 m_{H_u}^2 | H_u |^2 + m_{H_d}^2 | H_d|^2 + m_{S}^2| S |^2 \non \\
  &+& m_{{\tilde Q}_3}^2|{\tilde Q}_3^2| + m_{\tilde t_R}^2 |{\tilde t}_R^2|
 +  m_{\tilde b_R}^2|{\tilde b}_R^2| +m_{{\tilde L}_3}^2|{\tilde L}_3^2| +
 m_{\tilde  \tau_R}^2|{\tilde \tau}_R^2|\; .
\eeq
And the trilinear soft SUSY breaking interactions between the sfermion and Higgs fields are, 
\beq 
\label{Trimass}
-{\cal L}_{\mathrm{tril}}=  \lambda A_\lambda H_u H_d S + \frac{1}{3}
\kappa  A_\kappa S^3 + h_t A_t \tilde Q_3 H_u \tilde t_R^c - h_b A_b
\tilde Q_3 H_d \tilde b_R^c - h_\tau A_\tau \tilde L_3 H_d \tilde \tau_R^c
+ \mathrm{h.c.}\;.
\eeq
We work in the unconstrained NMSSM with non--universal soft terms at
the GUT scale. The three SUSY breaking masses squared for $H_u$, $H_d$
and $S$ which appear in ${\cal L}_{\mathrm{mass}}$ can be expressed
through their VEVs  by exploiting the three minimisation
conditions of the scalar potential.  While the MSSM Higgs sector at
tree-level can be described by only two free parameters (in general chosen to be
the mass of the pseudoscalar Higgs boson and $\tan\beta$), the Higgs sector 
of the NMSSM is parameterised by the six parameters
\beq
\lambda\ , \ \kappa\ , \ A_{\lambda} \ , \ A_{\kappa}, \ 
\tan \beta =\ \langle H_u \rangle / \langle H_d \rangle \ \mathrm{and}
\ \mu_\mathrm{eff} = \lambda \langle S \rangle\; .
\eeq
The brackets denote the VEV of the respective field inside. The sign conventions
are chosen such that $\lambda$ and $\tan\beta$ 
are positive, whereas $\kappa$,  $A_\lambda$, $A_{\kappa}$ and
$\mu_{\mathrm{eff}}$ can have both signs. \sn 

The Higgs sector consists of 3 CP-even Higgs bosons $H_i$ ($i=1,2,3$),
two CP-odd states $A_j$ ($j=1,2$) and two charged Higgs scalars
$H^\pm$. The neutral Higgs bosons are ordered by ascending mass with
$H_1$ ($A_1$) being the lightest CP-even (odd) Higgs boson. 
As higher order corrections to the Higgs sector are important and have
to be considered in order to calculate the Higgs sector as accurately
as possible, also the parameters from the non-Higgs sector, which enter 
through the loop corrections, have to be specified. These are the soft SUSY breaking mass
terms in Eq.~(\ref{Lmass}) for the scalars as well as the trilinear
couplings in Eq.~(\ref{Trimass}) and the gaugino soft SUSY breaking
mass parameters given by
\beq 
-{\cal L}_\mathrm{gauginos}= \frac{1}{2} \bigg[ M_1 \tilde{B}  
\tilde{B}+M_2 \sum_{a=1}^3 \tilde{W}^a \tilde{W}_a +
M_3 \sum_{a=1}^8 \tilde{G}^a \tilde{G}_a  \ + \ {\rm h.c.} 
\bigg].
\eeq

\section{The Scan \label{sec:scan}}
In the following we will perform a scan in the NMSSM parameter space
in order to investigate the Higgs sector in view of the recent LHC
Higgs search results together with the resulting possible theoretical
and phenomenological implications. When performing our scan we seek to
generate a Higgs spectrum where one of the scalar Higgs bosons
corresponds to a state with mass value around 126~GeV leading to event
rates in its production which are compatible with the LHC results. We
furthermore keep the fine-tuning 
\cite{King:2012is,bbsuppressandfinetune,SusyFineTuning,NMSSMtuning1}
as low as possible by demanding light top squark masses and/or small
mixing in the stop sector. \sn

For the calculation of the SUSY particle and NMSSM Higgs boson
spectrum and branching ratios we use the program package {\tt
  NMSSMTools} \cite{nmhdecay1,nmssmtools}.
The higher order corrections to the NMSSM Higgs boson 
masses are important \cite{effpotnmssm}
and have been included in {\tt
  NMSSMTools} up to ${\cal O}(\alpha_t \alpha_s + \alpha_b \alpha_s)$
for vanishing external momentum. Within the package the Fortran code
{\tt NMHDECAY} \cite{nmhdecay1}, an NMSSM extension of the Fortran code {\tt
  HDECAY} \cite{hdecay,susyhit}, provides the Higgs decay widths and branching
ratios, while the SUSY particle branching ratios are obtained from the
Fortran code {\tt NMSDECAY} \cite{nmsdecay} based on the generalisation of the
Fortran code {\tt SDECAY} \cite{susyhit,sdecay} to the NMSSM particle spectrum. The
output of the NMSSM particle spectrum, mixing angles, decay widths and
branching ratios is provided in the SUSY Les Houches Accord (SLHA) format
\cite{slha}. Being interfaced with {\tt micrOMEGAs} \cite{micromegas}, also the
relic abundance of the lightest neutralino $\tilde{\chi}_1^0$ as
the NMSSM Dark Matter candidate can be evaluated with {\tt
  NMSSMTools}. Furthermore, the package checks for the constraints
from low-energy observables as well as from Tevatron and LEP. For
details, we refer the reader to the program webpage
\cite{nmssmtools}.\footnote{Concerning the value of $g-2$, it is
  non-trivial to find
  parameter combinations which can explain the 2$\sigma$ deviation from
  the SM value. In our analysis we do not further consider this constraint.} \sn 

In order to restrict the parameter range for our scan we
are guidelined by the following objectives which follow from theoretical and experimental
considerations: 
\begin{itemize}
\item To keep the Higgs mass corrections (governed by the corrections
  from the (s)top sector) and hence the amount of fine-tuning as low as
 possible, the tree-level mass of the lightest Higgs boson is
  maximized by fixing $\tan\beta$ to small values chosen as
\beq
\tan\beta=2, \;4 \;.
\eeq
\item Also the effective $\mu_{\mathrm{eff}}$ parameter is kept as low
  as possible in order to avoid fine-tuning. It is varied in
  the range 
\beq
100 \; \mbox{GeV} \le \mu_{\mathrm{eff}} \le 200 \; \mbox{GeV} \; .
\eeq
Although we did not further consider the constraint coming from the
anomalous magnetic moment of the muon, we decided to take positive
values of $\mu_{\mathrm{eff}}$ as, similarly to the MSSM $\mu$
parameter, positive values are favoured when this constraint is
included, see {\it e.g.}~\cite{bbsupprandloop2}.
\item We shall be interested exclusively in large values of $\lambda$
  in order to increase the tree-level mass of the CP-even Higgs boson
  associated with the 126~GeV Higgs boson resonance. At the same time we pay
  attention that it remains small enough to ensure the validity of
  perturbation theory up to large scales, chosen to be the GUT scale
  here. This also constrains possible values of $\kappa$. Based on the
  results from the two-loop renormalisation group running down to
  1~TeV with and without the possibility of exotic extra matter \cite{King:2012is} we
  hence perform our scan in the ranges 
\beq
0.55 \le \lambda \le 0.8 \quad \mbox{and} \quad 10^{-4} \le \kappa \le
0.4 \;.
\eeq
\item The soft SUSY breaking trilinear couplings $A_\lambda$ and
  $A_\kappa$ are varied in the ranges 
\beq 
-500\; \mbox{ GeV} \le A_\kappa \le 0 \; \mbox{GeV} \quad \mbox{and}  \quad
200 \; \mbox{GeV} \le A_\lambda \le 800 \; \mbox{GeV} \; .
\eeq
\item For fine-tuning reasons we keep the soft SUSY breaking masses of the
  stop sector rather low and vary them simultaneously as
\beq
500 \; \mathrm{ GeV} \le M_{\tilde{Q}_{3}} = M_{\tilde{t}_R} \le 800\;
\mathrm{ GeV} \;.
\eeq
For $A_U$ ($U\equiv u,c,t$)\footnote{In {\tt NMSSMTools} there is no
  distinction between $A_u, A_c, A_t$.} we choose two representative
values corresponding to low and large mixing, 
\beq 
A_U= 0 \; \mbox{GeV} \quad \mbox{and} \quad 1 \; \mbox{TeV} \; . 
\eeq
Our lightest stop mass is hence about 400~GeV and in accordance with the
  LHC constraints \cite{stopsearchx}.\footnote{In scenarios with a
    very small mass difference between the lightest stop $\tilde{t}_1$ and the lightest
    neutralino $\tilde{\chi}_1^0$ assumed to be the lightest SUSY
    particle, stop masses down to about $100$-$130$~GeV are still
    allowed for $m_{\tilde{\chi}_1^0} \ge 90$~GeV
    \cite{lightstop1,lightstop2}.}
\item In order to comply with the present LHC search
  bounds \cite{squarksearch}, we conservatively set the soft SUSY
  breaking masses of the squark sector of the first two generations
  equal to 2.5 TeV and, for simplicity, also those of the slepton 
  sector apart from the soft SUSY breaking stau masses. The latter are chosen
  equal to 300~GeV. This way we still allow for rather light stau
  masses but are conservative enough to fulfill the latest LHC results
  \cite{stausearch}. It should be noted, however, that our results
  almost do not change by choosing different values in the stau
  sector\footnote{Also the SUSY breaking masses of the squarks of the first
    two generations barely influence the outcome of the scans.} as the influence
  of the slepton sector on the Higgs mass corrections is
  negligible. And contrary to the MSSM, light stau masses here do not
  lead to an enhancement of the partial width into photons
  \cite{carenastau}, as we have chosen small values of $\tan\beta$ and
  $\mu_{\mathrm{eff}}$. We
  furthermore set the trilinear couplings of the down and lepton
  sector equal to 1~TeV and the right-handed soft  
  SUSY breaking sbottom mass equal to 2.5~TeV. This results in light sbottom masses of about 
 $500 \mbox{ GeV} \lsim m_{\tilde{b}_1} \lsim 800 \mbox{ GeV}$. 
  Hence we have ($D\equiv d,s,b$, $E\equiv
  e,\mu,\tau$) 
\beq
M_{\tilde{u}_R}&\!\!\!=\!\!\!& M_{\tilde{c}_R} = 
M_{\tilde{D}_R} =M_{\tilde{Q}_{1,2}}= 
M_{\tilde{e}_R}  = M_{\tilde{\mu}_R}  = M_{\tilde{L}_{1,2}}=
2.5 \; \mbox{TeV},
\nonumber \\ 
M_{\tilde{\tau}_R} &\!\!\!=\!\!\!& M_{\tilde{L}_3} = 300\; \mbox{GeV} \;, \quad
 A_{D} =  A_E = 1 \; \mbox{TeV}  \;.
\eeq
\item The gluino soft SUSY breaking mass parameter has been set
  to 
\beq 
M_3 = 1 \; \mbox{TeV} \; . 
\eeq
The remaining two soft SUSY breaking gaugino 
  parameters have been chosen $M_1=150$~GeV and $M_2=300$~GeV.
\end{itemize}
It should be noted that in {\tt NMSSMTools} the NMSSM-specific input
parameters $\lambda, \kappa, 
A_\lambda$ and $A_\kappa$ according to the SLHA format are understood
as running $\overline{\mbox{DR}}$ parameters taken at the SUSY scale
$\tilde{M}=1$~TeV, while $\tan\beta$ is taken at the mass of the $Z$
boson, $M_Z$.  \sn
 
We remark, that at the cost of a more time consuming scan we could have enlarged our
parameter ranges of $A_\kappa$, $A_\lambda$ and $\kappa$. 
As will be evident from our numerical analysis later, the limitation
of the scan to this restricted parameter area nevertheless leads to a substantial
amount of parameter points which are compatible with the applied
constraints due to experimental results and fine-tuning
arguments. Note also, that choosing large positive values for
$A_\kappa$ for negative $\kappa$  leads to non self-consistent
solutions. Concerning $A_\lambda$, it is related to the charged Higgs
boson mass, which is below the experimental limit if $A_\lambda$ is
chosen too small. A posteriori it also turned out that the chosen upper bound
of $A_\lambda$ was largely sufficient to capture the maximum of
allowed parameter points which can be achieved for the chosen
$A_\kappa$ range. \sn

The parameter scan is further restricted by demanding the NMSSM Higgs
spectrum to fulfill the following conditions:
\begin{itemize}
\item We demand one of the scalar Higgs bosons, which we will denote from
  here on by $h$, to have its mass in the range 
\beq 
\underline{\mbox{scalar Higgs boson $h$:}}  \quad 
124 \, \mathrm{ GeV} \le m_h \le 127 \, \mathrm{ GeV} \;,
\label{eq:cond1}
\eeq
 where we have
 conservatively assumed a $3\sigma$ error on the mass value of the
 scalar particle discovered at the LHC \cite{:2012gk,:2012gu}.
\item In order to explore the possibility of an enhanced branching
  ratio into photons, we furthermore demand that the
  $\gamma\gamma$ rate around the invariant mass value 126~GeV fulfills:
\beq
  \mbox{rate for the } \gamma\gamma \mbox{ final state
  normalised to the SM value} \gsim 0.8 \;.
\label{eq:cond2}
\eeq 
\item We do not put any restrictions on the rates in the massive
  gauge boson and fermion final states.
\item For the other Higgs bosons, {\it i.e.}~the pseudoscalar Higgs
  bosons and the scalar Higgs bosons outside the mass range around
  126~GeV, we check if they have not been
  excluded by the LEP, Tevatron and LHC searches. Otherwise the whole parameter point is
  rejected. We have taken into account the newest exclusion limits in
  the various final states reported by the experiments
  \cite{higgsexclusionagamgamzz,higgsexclusioncgamgam,awwkyoto,abbkyoto,atautaukyoto,cwwkyoto,czzkyoto,cbbkyoto,ctautaukyoto}, which we have implemented in {\tt NMSSMTools}.
\end{itemize}

Our choice of parameters results in rather low top squark, charged
Higgs boson and chargino masses, still compatible, however, with present LHC SUSY
search results. For $\tan\beta=2$ we have
\beq
m_{\tilde{t}_1} &=& 400-820 \mbox{ GeV}\, , \quad
m_{\tilde{t}_2} = 530-890 \mbox{ GeV} \, , 
\\
M_{H^\pm} &=& 200-500 \mbox{ GeV}\, ,  \quad M_{\tilde{\chi}_1^\pm} =
105-165 \mbox{ GeV} \, , \quad
M_{\tilde{\chi}_2^\pm} = 345-360 \mbox{ GeV}\; ,
\eeq
and similar values for $\tan\beta=4$.
The stop mass values are small enough so that the fine-tuning is 
expected to be rather low. \sn

 We finally remark that we did not restrict our parameter points
 taking into account the relic density. We checked, however, that
 there is a substantial amount of
 parameter points which lead to relic densities due to a neutralino DM
 candidate, which are smaller than the WMAP
 value. To achieve the correct amount of relic density another
 candidate than the neutralino would have to be thought
 of. Furthermore, we convinced ourselves that {\it e.g.} by slightly changing the
 values of the gaugino mass parameters $M_1$, $M_2$ the correct amount
 of relic density could be achieved, while the Higgs mass spectrum
 remains practically unchanged, so that we did not further
 consider this constraint. For discussions taking into account DM
 constraints, see {\it
   e.g.}~\cite{constrainednmssm,comparative,relicdensity1,relicdensity2,relicdensity}.

\section{NMSSM Higgs boson production and decay \label{sec:proddec}}
In order to decide whether the 126~GeV NMSSM Higgs boson reproduces the
rates as measured by the experiments, its production cross sections
and branching ratios have to be investigated. In the following 
the dominant production process through gluon fusion
and its modification with respect to the SM will be discussed in detail. We furthermore
investigate the NMSSM Higgs branching ratio into photons, as the LHC 
experiments see a slight excess here with respect to the 
SM. With the presently available data, this has to be taken with due
caution, however, as it could still turn out to be a statistical
fluctuation. If it persists, however, it is a hint towards New
Physics and shall be taken into account in our analysis. We start
with some preliminary remarks and set up our notation. \sn

At the LHC, for small values of $\tan\beta$, the production processes
for a single neutral CP-even NMSSM Higgs boson $H_i$ ($i=1,2,3$) or a
CP-odd Higgs state $A_j$ ($j=1,2$) are given by
\vspace*{0.1cm}
\beq
\begin{array}{lll}
\hspace*{-0.5cm} \mbox{Gluon fusion: } & gg \to H_i \; \mbox{and} \; gg \to A_j &
\\[0.1cm]
\hspace*{-0.5cm} \mbox{Gauge boson fusion: } & qq \to qq + W^*W^*/Z^*Z^* \to qq H_i &
\\[0.1cm]
\hspace*{-0.5cm} \mbox{Higgs-strahlung: } & q\bar{q} \to Z^*/W^* \to H_i + Z/W &
\\[0.1cm]
\hspace*{-0.5cm} \mbox{Associated production with $t\bar{t}$: } & gg/q\bar{q} \to
t\bar{t} H_i \; \mbox{and} \; gg/q\bar{q} \to
t\bar{t} A_j &  \\[-0.3cm]
\phantom{x}
\end{array}
\eeq
where gluon fusion is the most important process followed by gauge
boson fusion. Higgs-strahlung and associated production with a top
quark pair\footnote{For small $\tan\beta$ values associated production
with a bottom quark pair is negligibly small.} only play a minor role
and are more important for the determination of Higgs boson
couplings. \sn

The NMSSM production processes and decay channels deviate from the
corresponding SM Higgs $H^{\scriptsize \mbox{SM}}$ processes due to modified Higgs couplings and
additional SUSY particles running in the loop mediated processes.
The couplings of the CP-even Higgs states $H_i$ (and also those of the
CP-odd states) depend on their decompositions
into the weak eigenstates $H_d$, $H_u$ and $S$, 
\beq
H_1 &=& S_{1,d}\ H_d + S_{1,u}\ H_u +S_{1,s}\ S\; ,\nonumber \\
H_2 &=& S_{2,d}\ H_d + S_{2,u}\ H_u +S_{2,s}\ S\; ,\label{eq:decomp} \\
H_3 &=& S_{3,d}\ H_d + S_{3,u}\ H_u +S_{3,s}\ S\; . \nonumber
\eeq
The coefficients $S_{i,u},S_{i,d}$ hence quantify the amount of up-
and down-likeness, respectively, while $S_{i,s}$ is a measure for the
singlet-component of a Higgs mass eigenstate.
Mixings between the $SU(2)$-doublet and singlet sectors are
proportional to $\lambda$, and can be sizeable for $\lambda \gsim 
0.3$, leading to significant effects on the NMSSM Higgs couplings and hence
phenomenology \cite{125singletdoublet,Ellwanger:2011aa,King:2012is,bbsuppressandfinetune}.
\sn

The inclusive production cross section $\sigma_{incl}$ for a CP-even
Higgs boson is composed of gluon
fusion, vector boson fusion, Higgs-strahlung and associated production
with $t\bar{t}$,
\beq
\sigma_{incl}(H) = \sigma (gg\to H) + \sigma (Hqq) + \sigma (WH) +
\sigma (ZH) +\sigma (t\bar{t}H) \approx  \sigma (gg\to H) \; , 
\eeq
with $H=H_i,H^{\scriptsize \mbox{SM}}$, respectively. It is dominated by the gluon fusion
cross section. For later convenience in the discussion of our results we normalise
the relevant quantities of the NMSSM Higgs bosons to the corresponding SM
counterparts. Thus we define the ratio  $R_{\sigma_{incl}}$ of the
NMSSM inclusive cross section compared to the SM one,
\beq
\label{incl}
R_{\sigma_{incl}}(H_i) \equiv \frac{\sigma_{incl} (H_i)}{\sigma_{incl} (H^{\scriptsize \mbox{SM}})}
\approx R_{\sigma_{gg}}(H_i) \; ,
\eeq
where we have used $R_{\sigma_{gg}}(H_i)$ defined as the ratio of the
NMSSM gluon fusion production cross section to the SM one, 
\beq  
\label{gg}
R_{\sigma_{gg}}(H_i)\equiv \frac{\sigma(gg\rightarrow
  H_i)}{\sigma(gg\rightarrow H^{\scriptsize \mbox{SM}})}
  \;.
\eeq
If not stated otherwise, in these and the following ratios the mass of the NMSSM Higgs boson
$H_i$ and the one of the SM Higgs $H^{\scriptsize \mbox{SM}}$ are taken to be the same and they
are subject to the constraint $M_{H^{\scriptsize \mbox{SM}}}=M_{H_i} \equiv m_h = 124-127$ GeV. \sn 

The ratio $R_{\Gamma_{tot}}$ for the total width compared to the SM Higgs
total width is given by
\beq 
\label{rattot}
R_{\Gamma_{tot}} (H_i)\equiv \frac{\Gamma_{tot}(H_i)}
{\Gamma_{tot}(H^{\scriptsize \mbox{SM}})} \;  .
\eeq
While in the SM the largest decay width of a Higgs boson of about
126~GeV is the one into $bb$, the most important search channels 
are given by the $\gamma\gamma$, the massive gauge boson and the
$\tau\tau$ final states. We define the ratios of the NMSSM Higgs
decay partial widths relative to the SM as ($ X=\gamma,W,Z,b,\tau$)
\beq
R_{\Gamma_{XX}} (H_i) \equiv \frac{\Gamma (H_i \to XX)}{\Gamma (H^{\scriptsize \mbox{SM}}
  \to XX)}  \;.
\eeq
The ratios of branching ratios are given by
\beq
R_{XX}^{BR} (H_i) \equiv \frac{BR (H_i \to XX)}{BR (H^{\scriptsize \mbox{SM}} \to XX)} =
\frac{R_{\Gamma_{XX}} (H_i)}{R_{\Gamma_{tot}} (H_i)} \;. 
\eeq
The experimentally observed rate in a given channel $X$ is given by the reduced cross
section $R_{XX}$ which is obtained from multiplying the Higgs production 
ratio relative to the SM, $R_{\sigma_{incl}}(H_i)$,
with the Higgs branching ratio for the channel of interest relative to the SM. For example,
for the two photon final state we have
\beq
R_{\gamma\gamma} (H_i) \equiv R_{\sigma_{incl}}(H_i)  R^{BR}_{\gamma\gamma}(H_i).
\label{eq:separate}
\eeq
The corresponding reduced cross sections in the other decay channels
$VV$ ($V=W,Z$), $bb$, $\tau \tau$ may be similarly expressed, namely:
\beq
R_{VV} (H_i) &\equiv& R_{\sigma_{incl}}(H_i) \, R^{BR}_{VV}(H_i) , \ \ \ \ 
R_{bb} (H_i) \equiv R_{\sigma(VH)}(H_i) \, R^{BR}_{bb}(H_i) ,
\nonumber \\ R_{\tau\tau} (H_i) &\equiv& R_{\sigma_{incl}}(H_i) \, R^{BR}_{\tau \tau}(H_i). 
\label{eq:redcxns}
\eeq
In the $bb$ final state we restrict ourselves to associated production
of the Higgs boson with a $W$ or $Z$ boson, as we will compare our
results later with values given for this channel by the experiments. \sn

It is important to note that there can be NMSSM spectra where
two neutral Higgs bosons lie close in mass. Due to the limited
experimental resolution these cannot be separated from each other and
both contribute to the signal. The program {\tt NMSSMTools} takes this
into account by super-imposing the signal from the nearby Higgs boson
with a Gaussian weighting. The width of the Gaussian smearing is
adapted to the respective experimental resolution in the different
final states, where clearly the $\gamma\gamma$ and $ZZ$ final states
have the best resolution, while the mass resolution in the $\tau\tau$
and $bb$ final states is less good, and in the $WW$ final state the mass cannot
be reconstructed. Hence, the ratios for the rates, $R_{XX}$, 
depending on the scenario and related
NMSSM spectrum under consideration, can be superpositions of rates of
different Higgs bosons. 
In favour of an unambiguous notation and to make contact with the
signal strengths $\mu=\sigma/\sigma_{\scriptsize \mbox{SM}}$ reported
by the LHC experiments, we denote by $\mu_{XX}$ the reduced cross sections
(\ref{eq:separate}), (\ref{eq:redcxns}), which are built up by the
superposition of the rates from the 126~GeV $h$ boson and another
Higgs boson $\Phi = H_i, A_j$, which is close by in mass, 
\beq
\mu_{XX} (h) \equiv R_{\sigma} (h) \, R_{XX}^{BR} (h) \;\; + \hspace*{-0.4cm}
\sum_{\scriptsize \begin{array}{c} \Phi\ne h \\ |M_{\Phi}\!-\!M_h| \le
  \delta \end{array}} \hspace*{-0.4cm} R_{\sigma} (\Phi)
\, R_{XX}^{BR} (\Phi) \, F(M_h, M_\Phi, d_{XX}) \;. \label{eq:mudef}
\eeq
Here $\sigma = \sigma(VH)$ in case $X=b$ and $\sigma = \sigma_{incl}$ otherwise.
By $\delta$ we denote the mass resolution in the respective $XX$ final
state and by $F(M_h,M_\phi,d_{XX})$ the Gaussian weighting
function as implemented in {\tt NMSSMTools}.
The experimental resolution of the different channels is taken
into account by the parameter $d_{XX}$, which influences the width
of the weighting function. We impose the restriction Eq.~(\ref{eq:cond2}) on
the thus calculated $\gamma\gamma$ rate, which in fact is the one
observed in experiment. Hence, in summary the 
conditions we impose on our parameter points are:
\beq
&&\hspace*{-1.6cm} 
\mbox{\underline{Conditions on the parameter scan:} } \nonumber \\[0.2cm]
&&\hspace*{-1.5cm}\begin{array}{ll}
\mbox{At least one CP-even Higgs boson $h$ with: } & 124 \mbox{ GeV }
\lsim M_{h} \lsim 127 \mbox{ GeV } \\[0.1cm]
\mbox{The reduced cross section for $\gamma\gamma$ must fulfill: } &
\mu_{\gamma\gamma} (h) \gsim 0.8 \quad \mbox{with } \\[0.2cm]
&
124 \mbox{ GeV } \!\lsim\! M_{h}= M_{H^{\scriptsize \mbox{SM}}} \!\lsim\! 127 \mbox{ GeV }
\end{array} 
\label{eq:cond}
\eeq
\vspace*{0.2cm}


\subsection{Higgs boson production through gluon
  fusion \label{sec:gluonfusion}}
\begin{figure}[b]
\vspace*{0.2cm}
\begin{center}
\includegraphics[width=10cm]{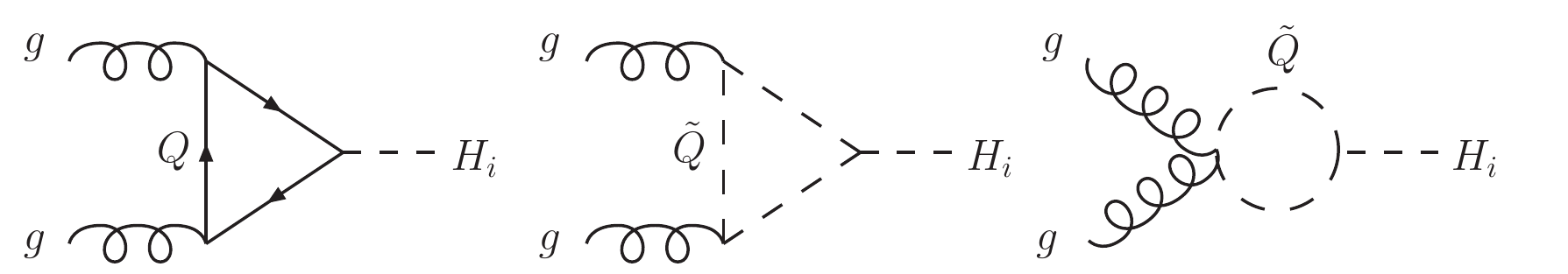}
\caption[ ]{\label{fig:nmssmggfus} Generic diagrams contributing gluon
  fusion production of $H_i $.}
\end{center}
\end{figure}

The cross section for NMSSM Higgs production via gluon fusion is
mediated by quark $Q$ and squark $\tilde{Q}$ triangle loops, {\it
  cf.}~Fig.~\ref{fig:nmssmggfus}. The latter become 
particularly important for squark masses below about 400 GeV
\cite{ggsquarkcontr,Djouadi:1998az,squarkandsuppr,fortschr}. At
leading order (LO) in the narrow-width approximation the hadronic
cross section for scalar Higgs bosons $H_i$ ($i=1,2,3$) can be cast into the
form \cite{fortschr,ggfus,qcdgaga}  
\beq 
\sigma_{LO} (pp \to H_i) &=& \sigma_0^{H_i}  \tau_{H_i} \frac{d{\cal
    L}^{gg}}{d\tau_{H_i}}
\\
\sigma_0^{H_i} &=& \frac{G_F \alpha_s^2 (\mu_R)}{288 \sqrt{2} \pi} \left|
  \sum_Q g^{H_i}_Q A_Q^{H_i} (\tau_Q) + \sum_{\tilde{Q}} g^{H_i}_{\tilde{Q}}
  A_{\tilde{Q}}^{H_i} (\tau_{\tilde{Q}}) \right|^2 \; ,
\eeq
with the gluon luminosity ${\cal L}_{gg}$, the Fermi constant $G_F$,
$\tau_{H_i} = M_{H_i}^2/s$, where $s$ denotes the squared  
hadronic c.m. energy and $\tau_X = 4 M_X^2/M_{H_i}^2$
($X=Q,\tilde{Q})$. The strong coupling constant $\alpha_s$ is taken 
at the scale $\mu_R$ chosen equal to the mass of
$H_i$. The form factors $A_{Q/\tilde{Q}}^{H_i}$ are given by
\beq
A_Q^{H_i} (\tau) &=& \frac{3}{2} \tau [1+(1-\tau) f(\tau)] \\
A_{\tilde{Q}}^{H_i} (\tau) &=& - \frac{3}{4} \tau [1-\tau f(\tau)] 
\eeq
and the function $f(\tau)$ reads
\begin{eqnarray}
f(\tau) & = & \left\{ \begin{array}{ll}
\displaystyle \arcsin^2 \frac{1}{\sqrt{\tau}} & \tau \ge 1 \\
\displaystyle - \frac{1}{4} \left[ \log \frac{1+\sqrt{1-\tau}}
{1-\sqrt{1-\tau}} - i\pi \right]^2 & \tau < 1 \; .
\end{array} \right.
\label{eq:ftau}
\end{eqnarray}
For large values of the loop particle masses the form factors become constant,
\beq
A_Q^{H_i} (\tau) &\to& 1 \qquad \mbox{for } M_{H_i}^2 \ll 4 m_Q^2 \\
A_{\tilde{Q}}^{H_i} (\tau) &\to& \frac{1}{4} \qquad \mbox{for }
M_{H_i}^2 \ll 4 m_{\tilde{Q}}^2 \;. 
\eeq
For small values of $\tan\beta$ the most important
contributions come from the top and stop loops. In order to
study the effect of the stop loops and their interplay with the
top quark loop, the Higgs couplings
to the top and stop quarks, $g^{H_i}_Q, g^{H_i}_{\tilde{Q}}$, have to
be investigated. Due to the diagonal 
gluon coupling to stops, in the loop only the Higgs couplings to two
equal stops can appear. Hence for $g^{H_i}_{\tilde{Q}}$ we have to
consider the couplings
\beq
g_{H_i \tilde{t}_1 \tilde{t}_1} &=& (S_{i,d} \cos\beta - S_{i,u}
\sin\beta) \frac{M_Z^2}{m_{\tilde{t}_1}^2}
(\frac{1}{2} \cos^2 \theta_{\tilde{t}} - \frac{2}{3} \sin^2 \theta_W \cos
2\theta_{\tilde{t}}) + \frac{m_t^2 S_{i,u}}{m_{\tilde{t}_1}^2 \sin\beta}\nonumber \\
&& + \frac{1}{2} \sin 2\theta_{\tilde{t}} \frac{m_t}{m_{\tilde{t}_1}^2
  \sin\beta} \left[-\mu_{\scriptsize \mbox{eff}} S_{i,d} + A_t S_{i,u}
  - \frac{\lambda v\cos\beta}{\sqrt{2}} S_{i,s} \right]  \label{eq:stop1coupl} \\
g_{H_i \tilde{t}_2 \tilde{t}_2} &=& (S_{i,d} \cos\beta - S_{i,u}
\sin\beta) \frac{M_Z^2}{m_{\tilde{t}_2}^2}
(\frac{1}{2} \sin^2 \theta_{\tilde{t}} + \frac{2}{3} \sin^2 \theta_W \cos
2\theta_{\tilde{t}}) + \frac{m_t^2 S_{i,u}}{m_{\tilde{t}_2}^2 \sin\beta}\nonumber \\
&& - \frac{1}{2} \sin 2\theta_{\tilde{t}} \frac{m_t}{m_{\tilde{t}_2}^2 \sin\beta}
\left[-\mu_{\scriptsize \mbox{eff}} S_{i,d} + 
  A_t S_{i,u} - \frac{\lambda v\cos\beta}{\sqrt{2}} S_{i,s} \right] \label{eq:stop2coupl}
\eeq
The Higgs mixing matrix elements $S_{i,x}$ ($x=d,u,s$) have been
defined in Eq.~(\ref{eq:decomp}). 
We are interested in NMSSM parameter scenarios with one of the CP-even
Higgs bosons having a mass around 126~GeV and production rates which
are not too far away from the corresponding SM rates in the various
final states in order to comply with the LHC Higgs search results.  
For the sake of simplicity in the investigation of the couplings we
therefore define the SM limit of the NMSSM, which is given by first performing the
MSSM limit, which is recovered by $\lambda,\kappa 
\to 0$ with $\kappa/\lambda$ constant and keeping the parameters
$\mu_{\mathrm{eff}},A_\lambda$ and $A_\kappa$ fixed. Within the MSSM
limit then the decoupling limit is performed. In the thus defined SM limit the mixing matrix elements
of $h$ become
\beq
S_{i,d} \to \cos\beta \qquad S_{i,u} \to \sin\beta \qquad S_{i,s} \to 0 
\;.
\label{eq:lim1}
\eeq
And we get for the couplings to the stops 
\beq
g^{\scriptsize \mbox{SM}}_{h\tilde{t}_1 \tilde{t}_1} &=&\cos2\beta
\frac{M_Z^2}{m_{\tilde{t}_1}^2} 
(\frac{1}{2} \cos^2 \theta_{\tilde{t}} - \frac{2}{3} \sin^2 \theta_W \cos
2\theta_{\tilde{t}}) + \frac{m_t^2}{m_{\tilde{t}_1}^2}\nonumber \\
&& + \frac{1}{2} \sin
2\theta_{\tilde{t}} \frac{m_t}{m_{\tilde{t}_1}^2\sin\beta} \left[-\mu_{\scriptsize \mbox{eff}} \cos\beta + A_t \sin\beta \right]  \label{eq:stopcoupl11}\\
g^{\scriptsize \mbox{SM}}_{h\tilde{t}_2 \tilde{t}_2} &=& \cos2\beta
\frac{M_Z^2}{m_{\tilde{t}_2}^2}  
(\frac{1}{2} \sin^2 \theta_{\tilde{t}} + \frac{2}{3} \sin^2 \theta_W \cos
2\theta_{\tilde{t}}) + \frac{m_t^2}{m_{\tilde{t}_2}^2 } \nonumber \\
&& - \frac{1}{2} \sin
2\theta_{\tilde{t}} \frac{m_t}{m_{\tilde{t}_2}^2 \sin\beta}
\left[-\mu_{\scriptsize \mbox{eff}} \cos\beta + A_t \sin\beta
\right] \label{eq:stopcoupl22}
\eeq
In the scenarios of the parameter scan which are left over after
applying our criteria Eq. (\ref{eq:cond}), for $\tan\beta=2$ the relations
(\ref{eq:lim1}) are approximately fulfilled apart from the singlet
component, which can take values of up to $\sim 0.1$. The
approximation gets worse in scenarios with strong singlet-doublet mixing,
where the singlet component can reach values of up to $\sim 0.6$. In
this case we can have suppressed couplings of the 126~GeV Higgs boson to
bottom quarks. For $\tan\beta=4$ the behaviour is similar for large mixing, while for
small mixing the deviations from this SM limit are more important. 
Note, that suppressed couplings to top quarks are
largely ruled out due to our demand of the $\gamma\gamma$ reduced
cross section
exceeding 80\% of the SM value. This can only be achieved if the
dominant production cross section through gluon fusion is large
enough, which is not the case for Higgs couplings to the top quarks
being too suppressed compared to the SM value. Nevertheless, also the
top quark couplings can be suppressed compared to the SM in the cases
where the branching ratio into $\gamma\gamma$ is enhanced or where 
the Higgs rates are built up by the superposition of rates
stemming from more than one Higgs boson, so that our restriction on
the $\gamma\gamma$ rate can be fulfilled. \s

The mixing angle $\theta_{\tilde{t}}$ which diagonalises the stop mass
matrix is given by
\beq
\sin 2\theta_{\tilde{t}} = \frac{2 m_t
  (A_t-\mu_{\scriptsize \mbox{eff}}/\tan\beta)}{m_{\tilde{t}_1}^2-m_{\tilde{t}_2}^2} \;,
\eeq
where $m_{\tilde{t}_{1(2)}}$ denotes the lighter (heavier) stop quark
mass. For the sake of the discussion we assume $\mu_{\scriptsize \mbox{eff}}$ to be
zero.\footnote{Our values of $\mu_{\scriptsize \mbox{eff}}$ are small enough not to
  change the conclusions for non-zero values.} With this approximation
and neglecting small $D$-term contributions we
have for $M_{\tilde{Q}_{3}} = M_{\tilde{t}_R}$ the mass 
difference $m_{\tilde{t}_2}^2 - m_{\tilde{t}_1}^2 = 2 m_t
A_t$. For large values of $A_t$, the $m_t^2$ term and the last term in
(\ref{eq:stopcoupl11}) therefore have opposite sign (in
(\ref{eq:stopcoupl22}) same sign). Neglecting the small $D$-term
contribution given by the first term in the coupling, the Higgs
coupling to the lighter stops $\tilde{t}_1$ becomes negative in this
case. Assuming $\tilde{t}_1$ to be 
relatively light, the contribution from the $\tilde{t}_1$ loop will be
more important and we will not consider the $\tilde{t}_2$ loop
contribution in this case. The $H_i$ coupling to the top quarks on the
other hand is given by
\beq
g^{H_i}_Q \equiv g_{H_i tt} = \frac{S_{i,u}}{\sin\beta} \;,
\eeq
which becomes in the SM limit for the 126~GeV Higgs boson
\beq
g_{htt} = 1 \;.
\eeq
Hence for large values of $A_t$ the $\tilde{t}_1$ and the top loop
contributions interfere destructively so that the gluon fusion cross
section decreases. For small values of $A_t$ there
is no mixing in the stop sector leading to a positive Higgs coupling
to $\tilde{t}_1$ and constructive interference, thus enhancing the
gluon fusion production cross section. For non-zero intermediate $A_t$
values, the last two contributions of (\ref{eq:stopcoupl11}) cancel
each other (the exact value of $A_t$ depends on the specific
parameter choices), and the stop contribution is small.\footnote{The
  influence of the stop loop contributions on gluon fusion 
  and decay into photon final states has been discussed in the context of the
MSSM in detail in \cite{Djouadi:1998az}.} 
The Higgs coupling values to the sbottoms are 
hardly influenced by a change in
$A_t$ which enters in the mixing matrix elements $S_{i,x}$
only through higher order corrections to the Higgs boson masses.  

\subsection{Higgs decay width into two photons}

\begin{figure}[t]
\begin{center}
\includegraphics[width=10cm]{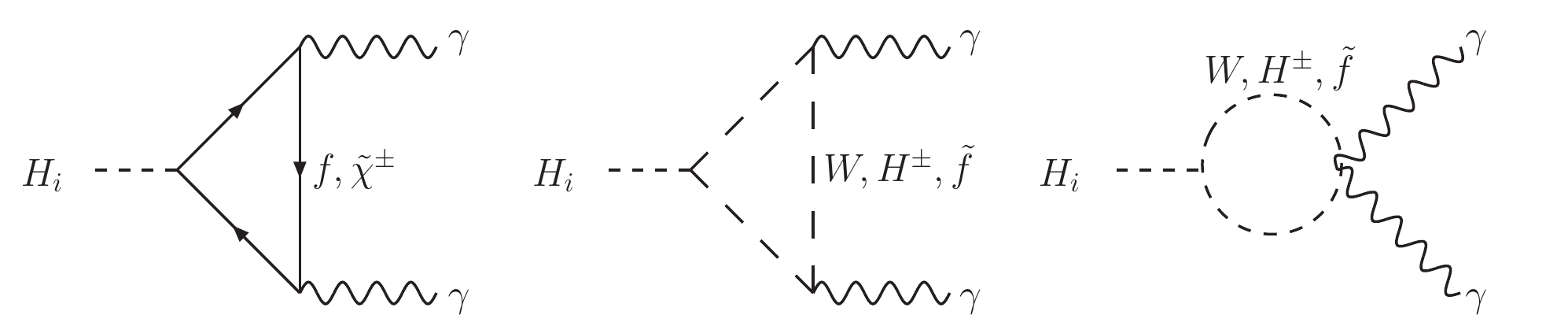}
\caption[ ]{\label{fg:mssmhgagalodia} Generic diagrams contributing to the decay
$H_i \to \gamma \gamma$.}
\end{center}
\end{figure}

The decays of the scalar NMSSM Higgs bosons into photons are mediated by $W$
boson and heavy
fermion loops as in the Standard Model and, in addition, by charged
Higgs boson, sfermion and chargino loops; the relevant diagrams are shown in
Fig.~\ref{fg:mssmhgagalodia}. The partial decay widths, adapted from
the MSSM result \cite{fortschr,qcdgaga,gagadecay}, are given by 
\begin{eqnarray}
\!\!\!\!\!\!
\Gamma (H_i\to \gamma\gamma )& = & \frac{G_{F} \alpha^{2}M_{H_i}^{3}}
{128\sqrt{2}\pi^{3}} \left| \sum_{f} N_{cf} e_f^2 g_f^{H_i}
A_f^{H_i}(\tau_f) + g^{H_i}_W A^{H_i}_W(\tau_W)
\right. \nonumber \\
& + & \left. g_{H^\pm}^{H_i} A_{H^\pm}^{H_i}(\tau_{H^\pm})
+ \sum_{\tilde \chi^\pm} g_{\tilde \chi^\pm}^{H_i}
A_{\tilde \chi^\pm}^{H_i} (\tau_{\tilde \chi^\pm}) +
\sum_{\tilde f}N_{cf}e_{\tilde f}^2 g_{\tilde f}^{H_i} A_{\tilde f}^{H_i}
(\tau_{\tilde f})
\right|^2 \; ,
\end{eqnarray}
with the colour factor $N_{cf}=1(3)$ for leptons (quarks) and $e_f$
denoting the electric charge of the loop particle. The form factors
are given by
\begin{eqnarray}
A_{f,\tilde \chi^\pm}^{H_i} (\tau) & = & 2 \tau \left[ 1+(1-\tau) f(\tau)
\right] \\ 
A_{H^\pm,\tilde f}^{H_i} (\tau) & = & - \tau \left[1-\tau f(\tau) \right] \\ 
A_W^{H_i} (\tau) & = & -\left[ 2+3\tau+3\tau (2-\tau) f(\tau) \right] \, ,
\end{eqnarray}
where $\tau = 4M_X^2/M_{H_i}^2$ with $M_X$ being the mass of the particle $X$ in the loop.
For large loop particle masses $M_X$ the form factors approach constant values,
\begin{eqnarray}
\begin{array}{llll}
A_{f,\tilde \chi^\pm}^{H_i} (\tau) & \to & \frac{4}{3} & \hspace*{1cm}
\mbox{for $M_{H_i}^2 \ll 4M_{f,\tilde \chi^\pm}^2$} \\[0.2cm]
A_{H^\pm,\widetilde{f}}^{H_i} (\tau) & \to & \frac{1}{3} & \hspace*{1cm}
\mbox{for $M_{H_i}^2 \ll 4M_{H^\pm,\widetilde{f}}^2$} \\[0.2cm]
A_W^{H_i} (\tau) & \to & - 7 & \hspace*{1cm} \mbox{for $M_{H_i}^2 \ll 4M_W^2$}
\, .
\end{array}
\end{eqnarray}
The Higgs couplings
to fermions, $W$ bosons, charged Higgs bosons and charginos,
appearing in the decay width into two photons, are given by
\begin{align}
g_f^{H_i} &= \left \{ \begin{array}{lll} S_{i,u}/\sin\beta & \mbox{ for
      } & f= \mbox{ up-type fermion} \\ S_{i,d} /\cos\beta & \mbox{
        for } & f= \mbox{ down-type fermion} \end{array}
  \right.  \\
g_W^{H_i} &= S_{i,d} \cos\beta + S_{i,u} \sin\beta 
\\
g^{H_i}_{\tilde{\chi}^\pm} &\equiv g^{H_i}_{\tilde{\chi}^\pm_k
  \tilde{\chi}^\mp_k} = \frac{2M_W}{M_{\tilde{\chi}_k^\pm}} \left[ q_{kk} S_{i,d} + s_{kk} S_{i,u} + r_{kk}
S_{i,s} \right] \displaybreak[3] \\  
g_{H^\pm}^{H_i} &= \frac{M_W^2}{M_{H^\pm}^2}\Bigg[\frac{\cos(2\theta_W)}{2\cos^2\theta_W}\Big(\cos^3\!\beta \;S_{i,d}+\sin^3\!\beta \;S_{i,u}\Big)\nonumber \displaybreak[3] \\
&+\frac{1}{2}\cos\beta \sin\beta\Big((3+\tan^2\!\theta_W)-4\lambda^2/g^2\Big)\big(\sin\!\beta \;S_{i,d}+\cos\!\beta\;S_{i,u}\big)\nonumber \displaybreak[3] \\
& +\frac{1}{\sqrt{2}g M_W}\Big(2\lambda\mu_{\text{eff}}+\sin2\beta(A_\lambda \lambda+2\kappa\mu_{\text{eff}})\Big)S_{i,s}\Bigg]\;.
\end{align}
The matrix elements $q_{kl},s_{kl},r_{kl}$ ($k,l=1,2$) in terms of the
matrix elements of the matrices $U,V$ diagonalising the chargino mass
matrix \cite{charginos} read
\beq
q_{kl} = \frac{1}{\sqrt{2}} U_{l2} V_{k1} \quad , \quad s_{kl} =
\frac{1}{\sqrt{2}} U_{l1} V_{k2} \quad , \quad r_{kl} =
\frac{\lambda v}{2\sqrt{2} M_W} U_{l2} V_{k2} \;.
\eeq
In the SM limit, defined in the previous subsection, the couplings become
\beq
g_f^{H_i} &\to& 1 \\
g_W^{H_i} &\to& 1 \\
g_{H^\pm}^{H_i} &\to& \frac{M_W^2}{M_{H^\pm}^2}\Bigg[\frac{\cos(2\theta_W)}{2\cos^2\theta_W}\Big(\cos^4\!\beta+\sin^4\!\beta\Big)+\cos^2\!\beta \sin^2\!\beta\Big((3+\tan^2\!\theta_W)-4\frac{\lambda^2}{g^2}\Big)\Bigg]\\
g^{H_i}_{\tilde{\chi}^\pm_k \tilde{\chi}^\pm_k} &\to& \frac{2 M_W}{M_{\tilde\chi_k^\pm}}\Big(q_{kk} \cos\beta + s_{kk} \sin\beta\Big)  \; .
\eeq
The sfermion loop contributions are only important for the
third generation with light masses. While it has been shown
in the context of the MSSM that the $\tilde{\tau}$ loop contributions
can enhance the decay width into photons \cite{carenastau}, this is
not the case here as we assume 
small values of $\tan\beta$ and $\mu_{\mathrm{eff}}$. Due to the small
$\tan\beta$ values also sbottom loops play a minor role. The most
important sfermion contribution comes from the stop loops in
particular a light $\tilde{t}_1$. The Higgs
coupling to the stops has been given in section \ref{sec:gluonfusion},
Eqs.~(\ref{eq:stop1coupl}), (\ref{eq:stop2coupl}) and for the SM limit in
Eqs.~(\ref{eq:stopcoupl11}), (\ref{eq:stopcoupl22}). 
Since in the Higgs decay into photons the quark and $W$
loop contributions interfere destructively, in the decay the
effect of the stop loops is opposite to the one in the production. For
$A_t=0$~GeV the stop loop contribution suppresses the decay
into photons, for $A_t=1$~TeV it leads to an enhancement.  
\sn

For light enough chargino and charged Higgs boson masses their loop
contribution also plays a role. We find that they can lead to an
enhancement for the partial decay width into photons (see also
\cite{chargedloops,bbsupprandloop3,bbsupprandloop4}). 

\section{Numerical Analysis \label{sec:numerics}}
In this section we show our numerical results.
When performing the scans we find scenarios in which both the lightest scalar Higgs
$H_1$ and the heavier one $H_2$ can have masses around 126~GeV. We
will call the respective Higgs boson in this case $h$. Not
all its couplings are necessarily equal or near the corresponding SM Higgs coupling
values. More importantly the reduced cross sections in the various final states, which
can be superpositions of signals from Higgs bosons with masses close
by, as defined in Eq.~(\ref{eq:mudef}), have to reproduce the
experimental results. To avoid a flood of plots, in the following we
will only show the ones for $\tan\beta=2$ and comment on the plots for $\tan\beta=4$.

\subsection{Parameter values and mass distributions}
\begin{figure}[t]
\includegraphics[width=8cm]{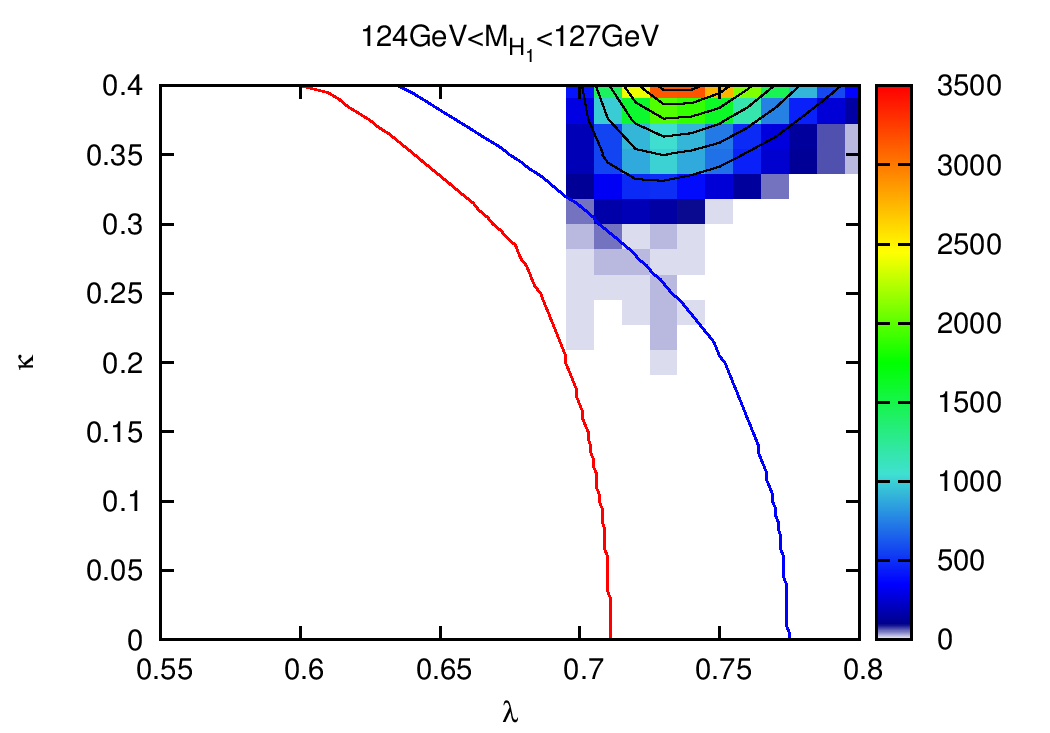}\includegraphics[width=8cm]{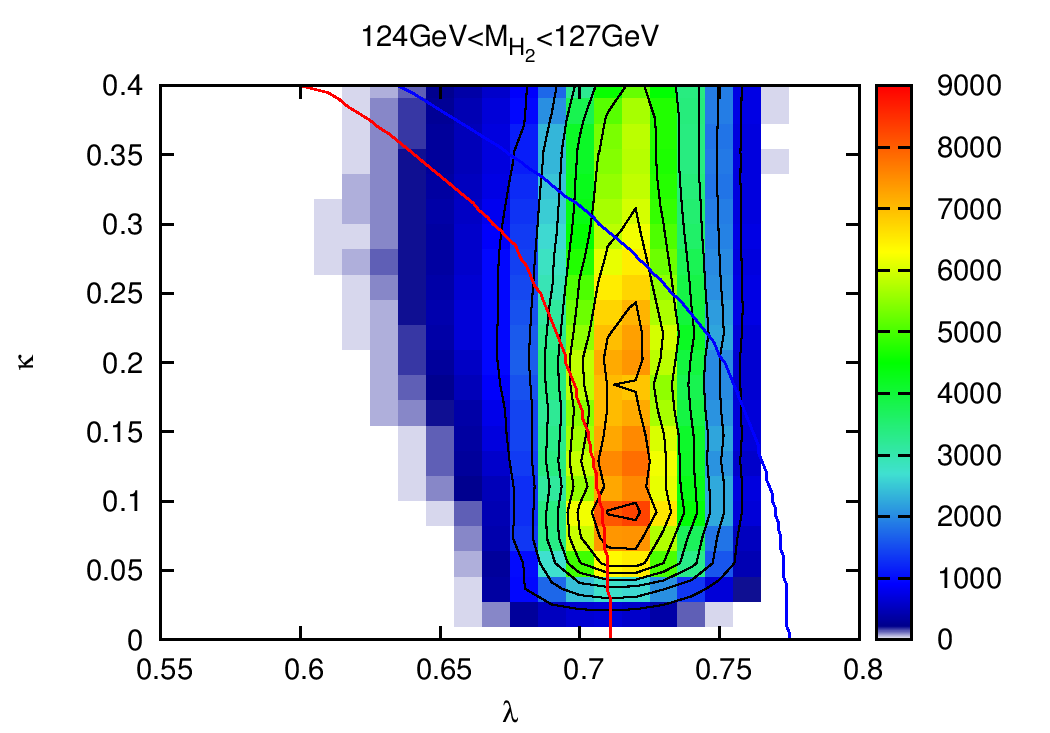}\\[0.2cm]
\includegraphics[width=8cm]{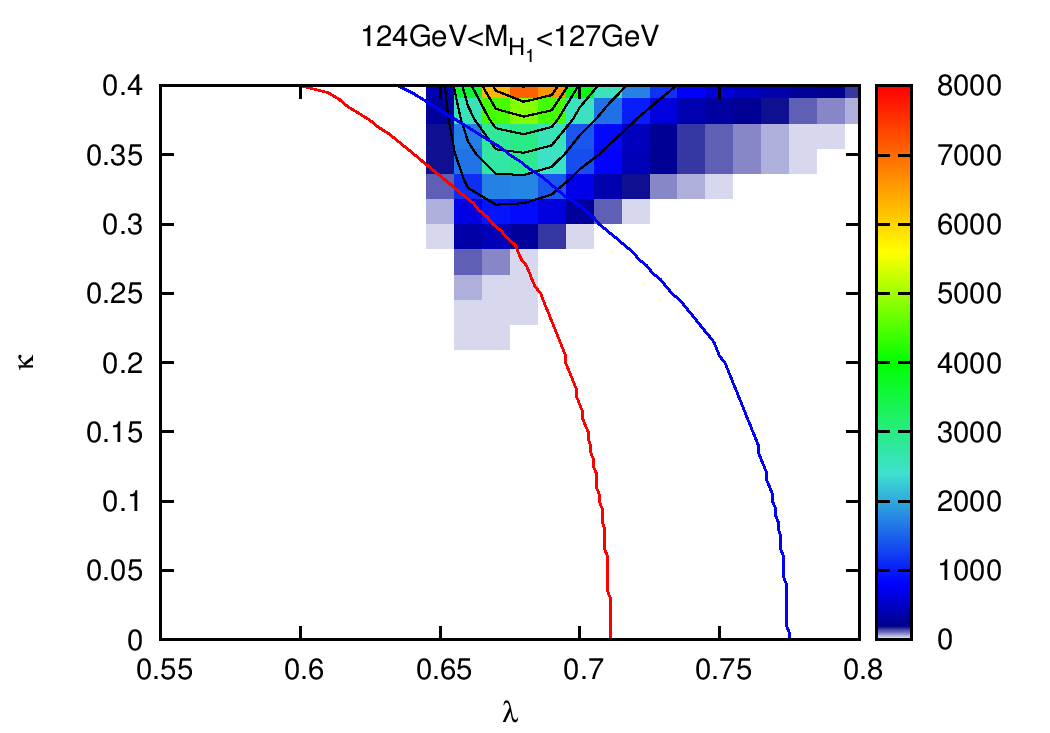}\includegraphics[width=8cm]{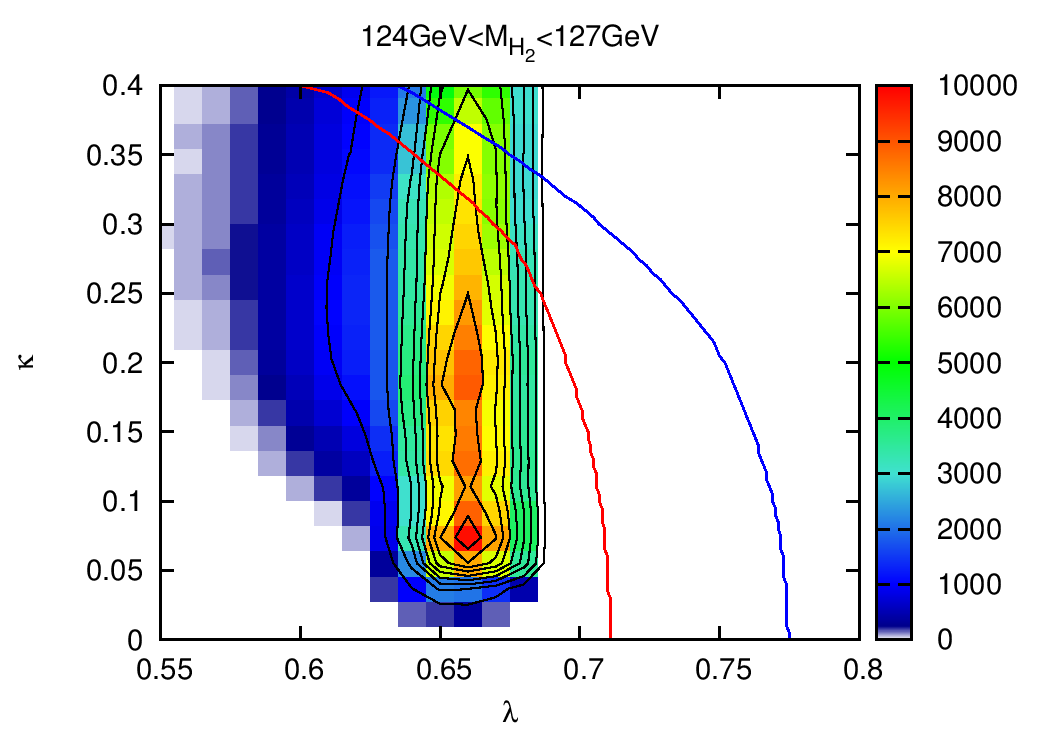}\\
\caption{The distribution of allowed parameter points in the
  $\kappa$-$\lambda$ plane for $h=H_1$ (left) and $h=H_2$ (right) for
  $A_t=0$~GeV (upper) and $A_t=1$~TeV
  (lower). The red (blue) contour lines show the two-loop upper bounds for
  $\lambda$ and $\kappa$ at 1~TeV in the NMSSM
  without (with) extra matter above 1~TeV. The colour code denotes the
  number of points.
\label{fig:distr1}}
\end{figure}
In Fig.~\ref{fig:distr1} we show the distributions of
the allowed parameter points in the $\kappa$--$\lambda$ plane leading to $H_1$
representing the CP-even Higgs boson with mass in the range 124--127~GeV
(left) and to $H_2$ (right) being $h$, respectively, 
for the two values of $A_t=0$~GeV and 1~TeV. The colour code denotes the number of points.
As can be inferred from the figures, we have much more points allowing for $h=H_2$ than
for $h=H_1$. This can be explained as follows. As we demand the stop
mass parameters to be rather low, which leads to smaller higher order corrections for
a fixed mixing, the parameter $\lambda$ at low energies must be large
enough to get to the right mass. The demand of perturbativity up to
the GUT scale then implies stringent constraints on the coupling
$\kappa$(1~TeV). In addition $\mu_{\scriptsize \mbox{eff}}$ is
required to be smaller than 200~GeV to avoid fine-tuning. Because the
masses of the extra NMSSM scalar and pseudoscalar states, which are
predominantly SM singlets, are set by $\kappa \,\mu_{\scriptsize
  \mbox{eff}}/\lambda$ these states tend to be lighter than the
126~GeV Higgs boson. As both the $H_1$ and $H_2$ mass values increase with
rising $\kappa$, for $H_1$ we need for the same reasons large $\kappa$
values of $\kappa \approx 0.4$,
while too large values of $\kappa$ lead to too large $H_2$ masses so
that here values $\kappa \approx 0.07$--$0.09$ are preferred. With increasing 
values of $A_t$ the stop mass corrections to the tree-level masses
become more important so that a 126~GeV Higgs mass can be attained
more easily and therefore more parameter points pass the
constraints. For the same reason the maximum of points is given for
smaller values of $\lambda$ now, decreasing from $\lambda \approx
0.73\, (0.72)$ at small stop mixing to $\lambda \approx 0.68\, (0.66)$ for
$h=H_1\, (H_2)$ at large mixing. \sn

\begin{figure}[t]
 \includegraphics[width=8cm]{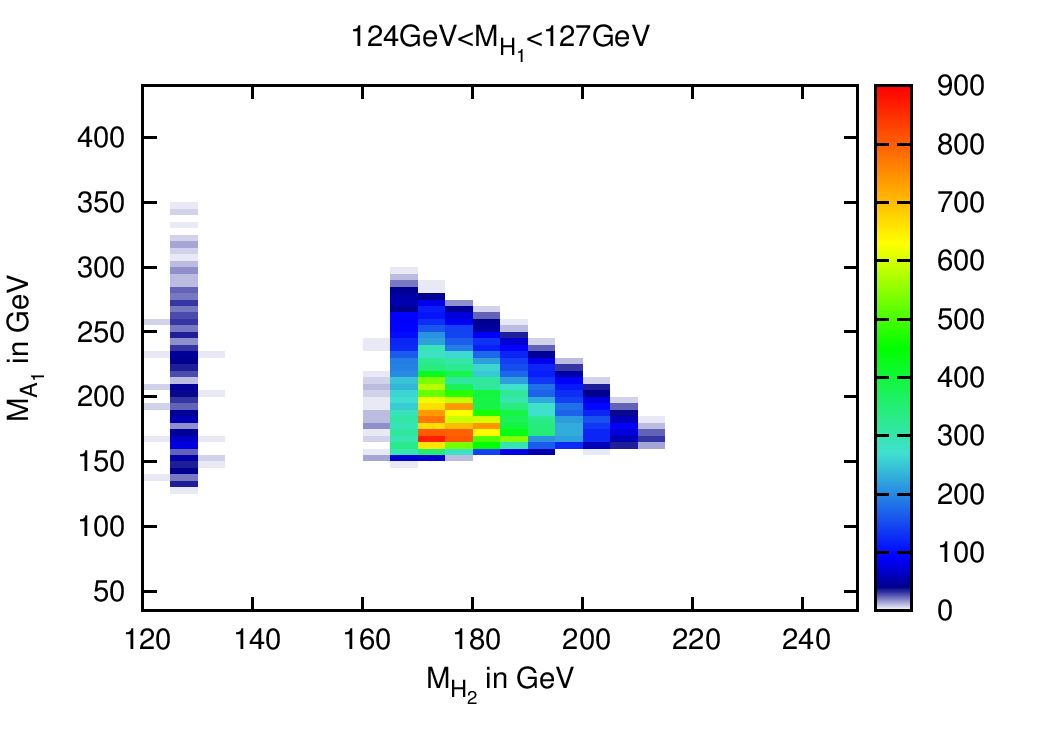}\includegraphics[width=8cm]{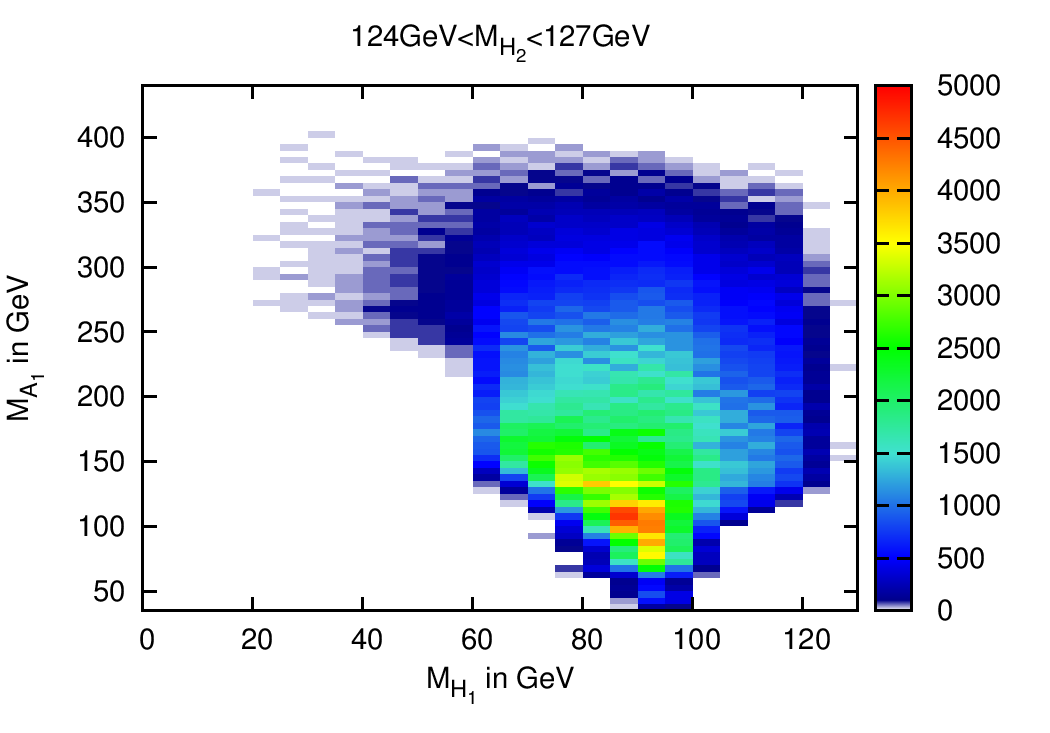}\\
 \includegraphics[width=8cm]{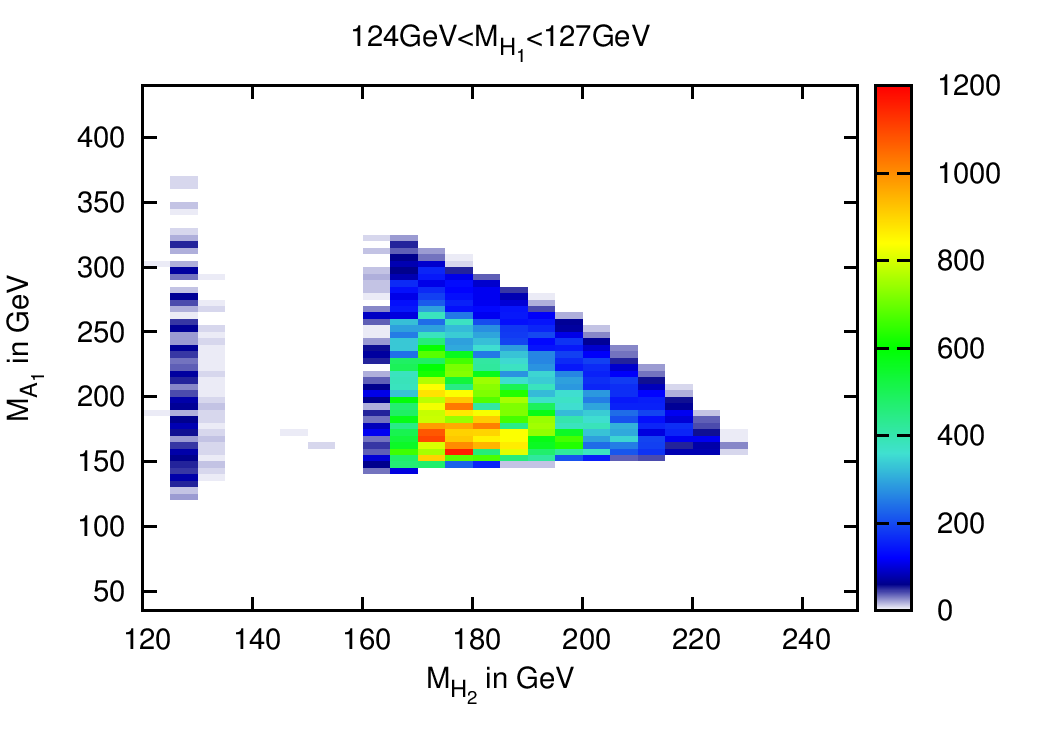}\includegraphics[width=8cm]{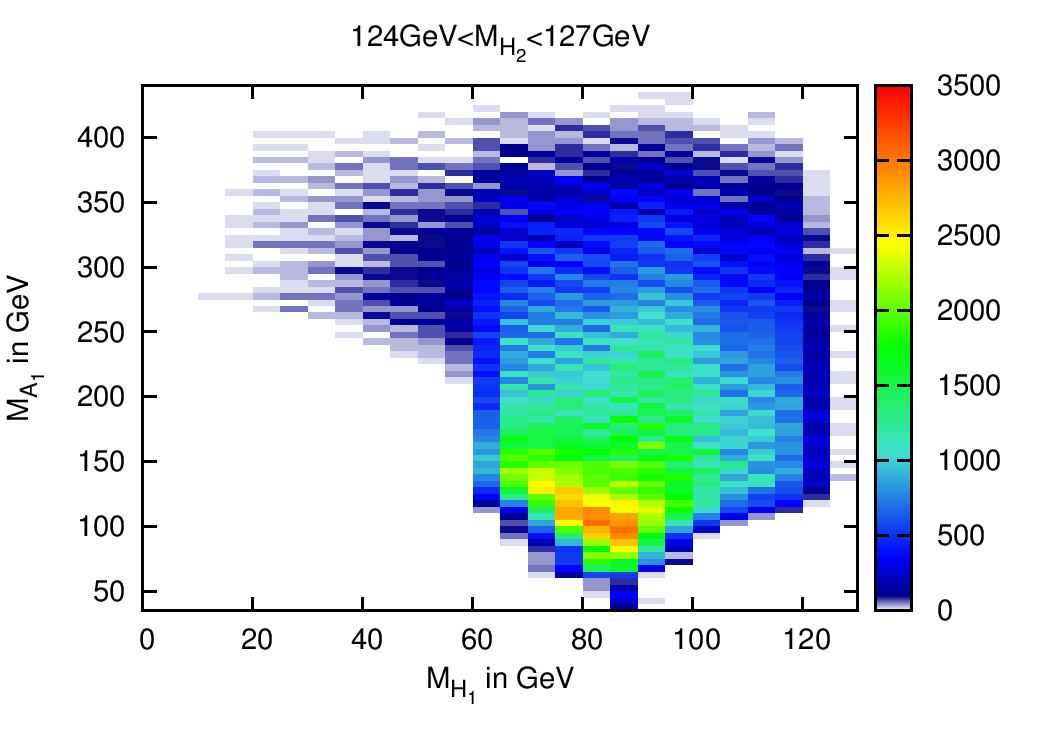}\\
\caption{Mass spectrum of $A_1$ and $H_2$ for $h=H_1$ (left) and of
  $A_1$ and $H_1$ for $h=H_2$ (right) with $A_t=0$~GeV (upper) and
  1~TeV (lower). The colour code denotes the number of points.}
\label{fig:massspectr}
\end{figure}
In the plots we also show the upper bounds on $\lambda$ and $\kappa$
imposed by perturbativity derived from the two-loop renormalisation group
running from the GUT scale down to 1~TeV. These limits can be somewhat
relaxed when allowing for extra exotic matter with mass around
1~TeV. They show that an $H_1$ Higgs boson with mass around 126~GeV can only be achieved
for large mixing with $A_t=1$~TeV. For lower values of $A_t$ even with
the inclusion of extra matter, this is not possible. The heavier Higgs boson
$H_2$ on the other hand can have a 126~GeV mass value with and
without exotic matter. We finally note 
that in case $A_t=0$~GeV, for $h=H_1$ the trilinear couplings
$A_\kappa,\, A_\lambda$ cluster around $(A_\kappa,A_\lambda)=(0\mbox{
  GeV},310 \mbox{ GeV})$ and for $h=H_2$ around $(A_\kappa,A_\lambda)=(-140\mbox{
  GeV},310 \mbox{ GeV})$. In case $A_t=1$~TeV, we have for $h=H_1$ the
maximum of points around  $(A_\kappa,A_\lambda)=(0\mbox{ GeV},340
\mbox{ GeV})$ and for $h=H_2$ around $(A_\kappa,A_\lambda)=(-140\mbox{
  GeV},340 \mbox{ GeV})$. 
\sn

Figure~\ref{fig:massspectr} shows the mass distributions of the
lighter neutral Higgs bosons for $H_1$ and $H_2$ being $h$,
respectively. For $h=H_1$ there exist parameter regions where $H_2$
and/or $A_1$ are very close in mass. Depending on the respective
experimental resolution in the investigated final state their signal
can superimpose the $h$ rate. This superposition has been taken into
account in the reduced cross sections discussed later. The maximum of
parameter points clusters around mass values $M_{H_2,A_1}\approx
(175,170)$~GeV. Also for $H_2$ 
with mass $\sim 126$~GeV the $H_1$ and/or $A_1$ state can be close in
mass and contribute to the signal. Their masses can be also much
smaller, however, so that $H_2$ decays into these final states become
possible, leading to distinct signatures \cite{ellpaper}. The
maximum parameter points are found for $M_{H_1,A_1} \approx
(85,110)$~GeV.  The masses of the heavier Higgs bosons $H_3$ and $A_2$ lie between
about 300 and 500 GeV.\sn

We remind the reader that in all plots we have already taken into account the
latest exclusion limits from LEP, Tevatron and LHC which apply to the non-$h$ Higgs
bosons. In particular for scenarios with $h=H_1$ this leads to a
substantial reduction of allowed parameter points. \sn

As for $\tan\beta=4$, it turns out that for small mixing no parameter
combination fulfills the conditions (\ref{eq:cond}) for the lightest
NMSSM Higgs boson $H_1$. Only for large
mixing a few hundred parameter points survive which cluster around
$(\kappa,\lambda)=(0.4,0.8)$.  For $h=H_2$ both in the low and in the
large mixing case the conditions are fulfilled with the maximum of
points in a somewhat extended region around
$(\kappa,\lambda)=(0.4,0.8)$. The reason is that for larger values of
$\tan\beta$ the tree-level upper mass bound is lower than for $\tan\beta=2$, so that more
substantial higher-order mass corrections are needed which in case $H_1$ is to have a
mass around 126~GeV can be achieved only for large mixing. If we now
apply the perturbativity bounds on $\kappa$ and $\lambda$ it turns out,
however, that none of the $h=H_1$ scenarios survives as even with
extra matter at 1~TeV the maximum allowed value is $\lambda=0.66$ for
$\kappa=0.4$. For the $h=H_2$ scenarios a few scenarios survive if
extra matter is included. Otherwise the perturbativity bounds imposing
a maximum value $\lambda=0.64$ for $\kappa=0.4$ are not respected. \sn

We summarise, that for $\tan\beta=2$ there are scenarios with $h=H_1$
which respect perturbativity in case of large mixing and inclusion of
extra matter at 1~TeV. For $h=H_2$ this is the case for both small and
large mixing and NMSSM with and without extra matter. For
$\tan\beta=4$ only $h=H_2$ scenarios survive and are compatible with
perturbativity for low and large mixing if extra matter is included.

\subsection{The \boldmath{$\gamma\gamma$} final state}
\begin{figure}[t]
 \includegraphics[width=8cm]{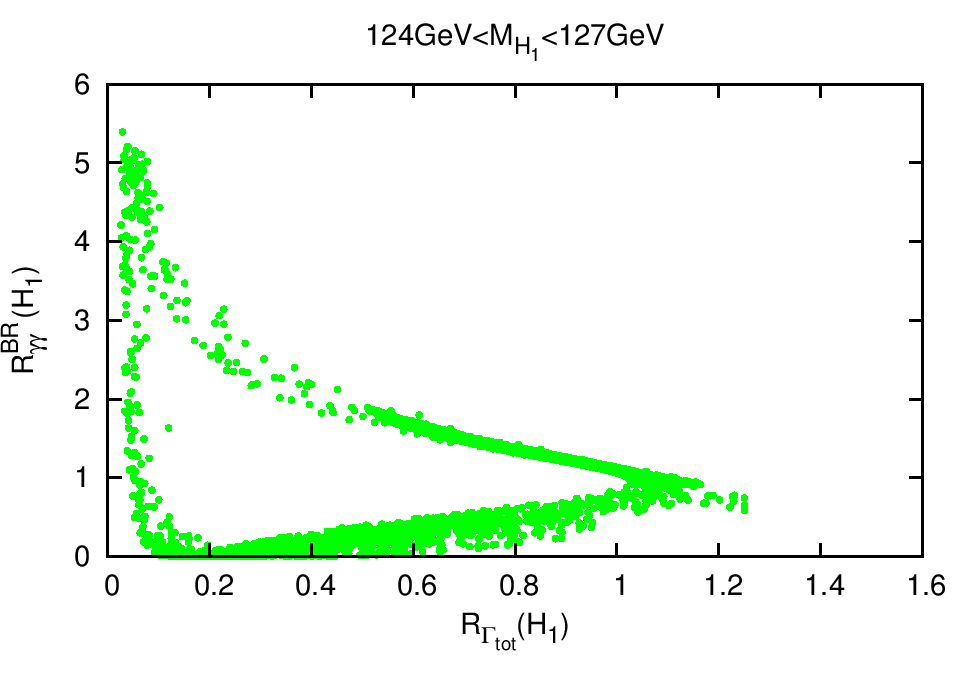}\includegraphics[width=8cm]{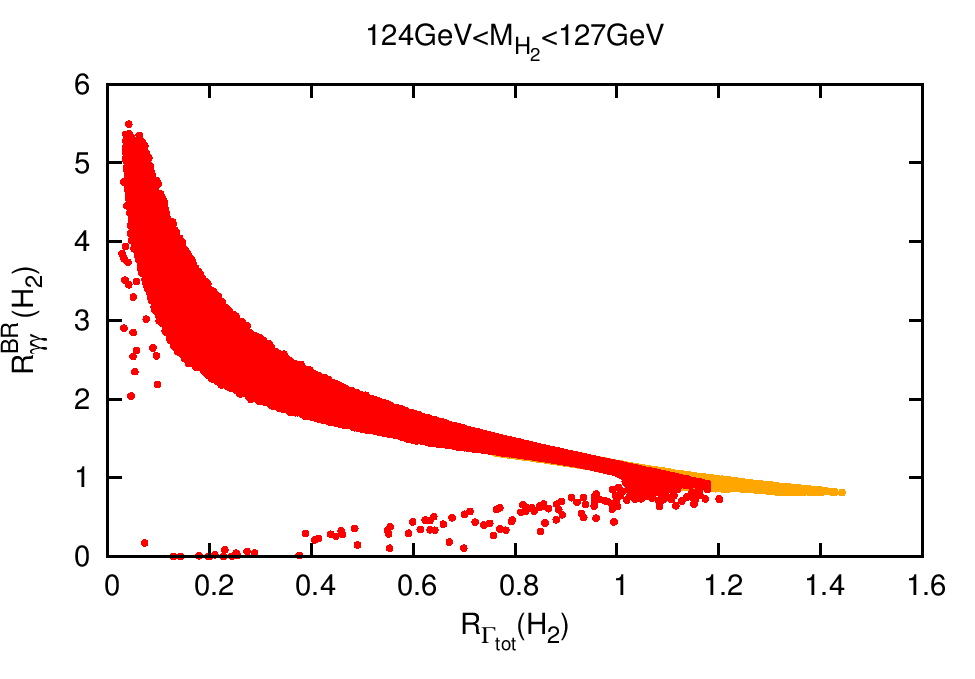}
\caption{Branching ratio into $\gamma\gamma$ relative to the SM
  against the normalised total width for $h=H_1$ (left) and
  $h=H_2$ (right) and $A_t=1$~TeV. The orange points (for $h=H_2$) indicate that
  non-SM decays are allowed.}
\label{fig:branratios}
\end{figure}
We first discuss the behaviour of the photonic branching ratio, which
is shown in Fig.~\ref{fig:branratios} with respect to the SM compared to
the normalised total width for $h=H_1$ and $H_2$, respectively, and
$A_t=1$~TeV. The plots for $A_t=0$~GeV look very similar and therefore
are not shown here. The branching ratio into $\gamma\gamma$ can be largely
enhanced up to $\sim 5.6$ times the SM value. This is due to a
substantially suppressed total width because of strong
singlet-doublet mixing of the Higgs boson with mass
$\sim 126$~GeV. Its coupling to bottom quarks is therefore strongly
reduced, leading to a small total width (dominated by the decay into
$b$-quarks) and hence an enhanced branching ratio. The increase in the
branching ratio, however, can also be due to an enhanced decay width into
photons caused by squark, charged Higgs and/or chargino loop
contributions, as has been discussed above. Therefore also for
enhanced total widths the branching ratio can be larger than in the SM
case. If, however, besides the couplings to bottom quarks also the other Higgs
couplings are substantially suppressed due to strong singlet-doublet mixing, the
loop-induced coupling to photons becomes very small, leading to small
branching ratios also in the case of small total widths.
For $h=H_2$ we can observe that the total width can be increased by up
to~$\sim 1.5$ compared to the SM. This happens where decays of $H_2$ into other lighter
Higgs bosons $H_1$ or $A_1$ and/or neutralino final states are
kinematically allowed \cite{ellpaper}. 
The relevant decays are
$H_2 \to \tilde{\chi}_1^0 \tilde{\chi}_1^0$, $H_2 \to H_1 H_1$ and
$H_2 \to A_1 A_1$, with the latter being rarely realised.\sn

\begin{figure}[t]
 \includegraphics[width=8cm]{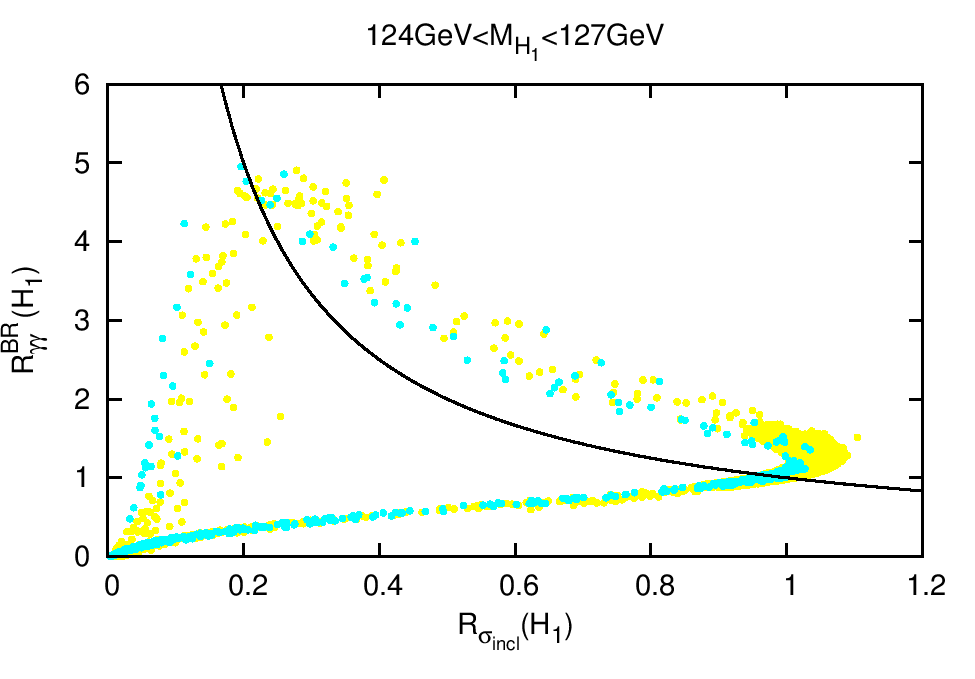}\includegraphics[width=8cm]{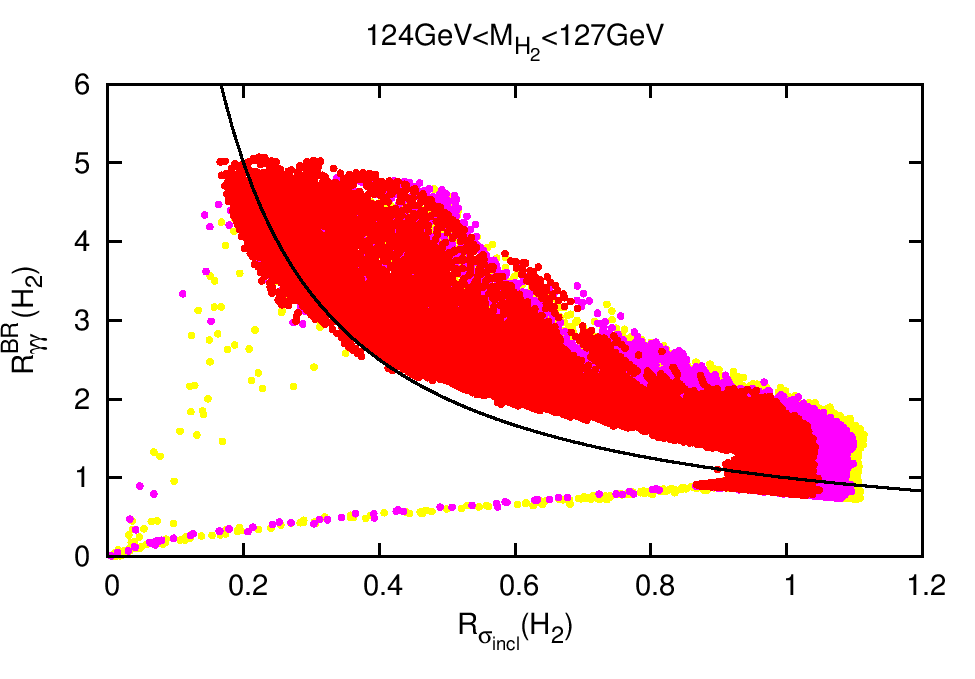} \\[0.2cm]
 \includegraphics[width=8cm]{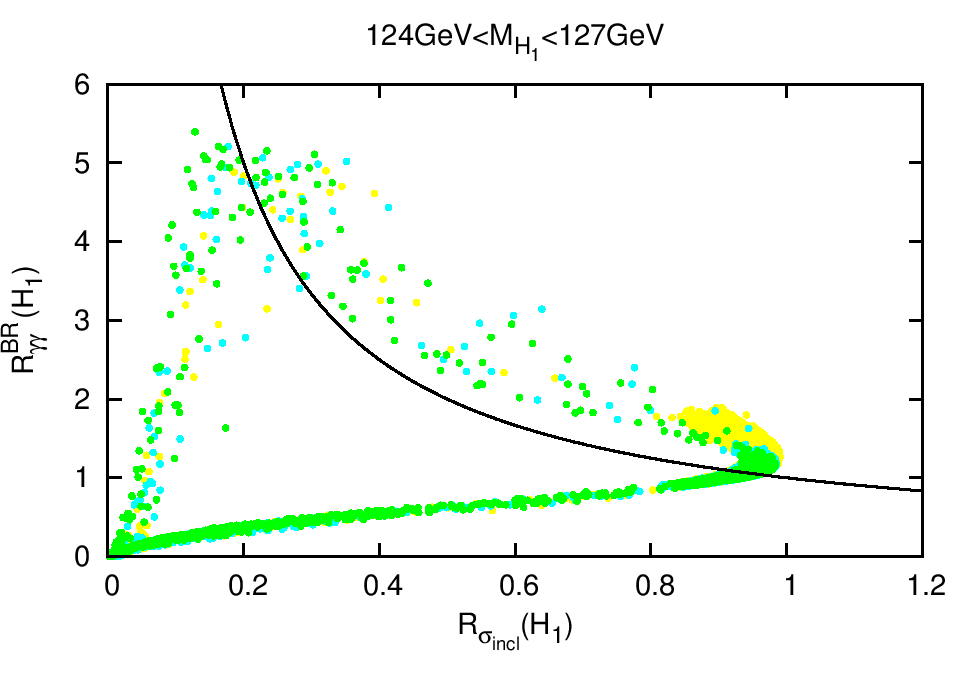}\includegraphics[width=8cm]{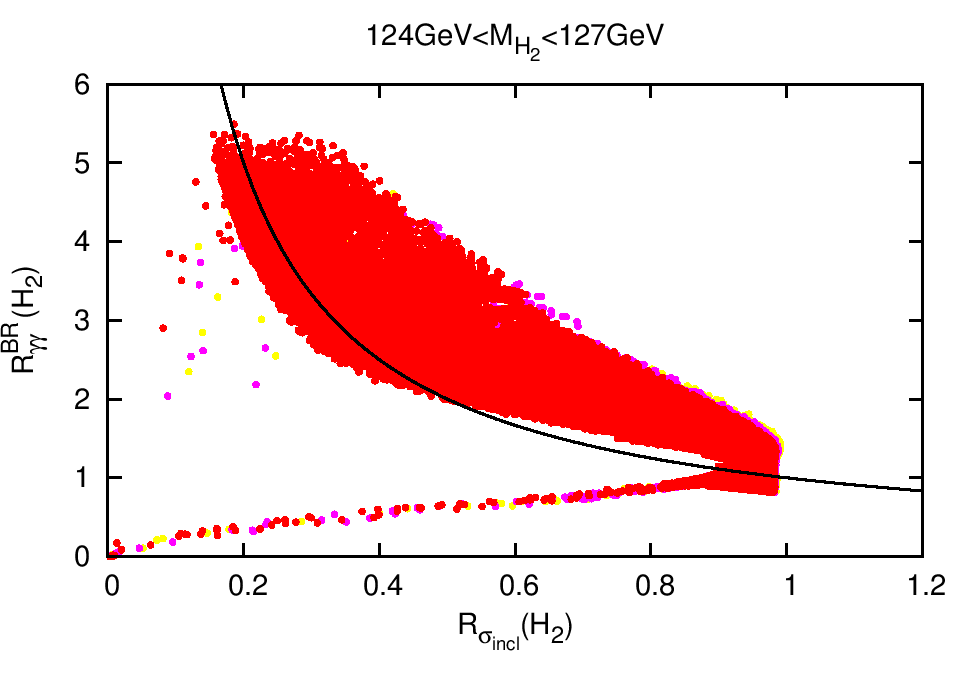}
\caption{Branching ratio into $\gamma\gamma$ relative to the SM against the 
inclusive production cross section relative to the SM for
$h=H_1$(left) and $h=H_2$(right) for $A_t=0$~GeV (upper)
and $A_t=1$~TeV (lower). For green/red points
the perturbation theory is valid up to the GUT scale. Cyan/pink points
require extra matter above $1\mbox{ TeV}$ and yellow points violate the
two-loop upper bounds on $\lambda, \kappa$ both with and without extra matter. 
Points above the black line lead to an enhanced $R_{\gamma\gamma}$.}
\label{fig:gamgamrate}
\end{figure}
The corresponding plots to Fig.~\ref{fig:branratios} for $\tan\beta=4$
show a similar behaviour with altogether less parameter points,
however, and a maximum photonic branching ratio enhancement of
$R_{\gamma\gamma} \approx 5$ for both $h=H_2$ and $H_1$ (with only the
large mixing case surviving here). And for the total width the maximum
value is $R_{\Gamma_{\scriptsize \mbox{tot}}} \approx 1.35$ due to
$H_2$ decays into light Higgs bosons or neutralinos. \sn

With Fig.~\ref{fig:gamgamrate} we discuss the interplay of production
and decay on the photon rate. We show the branching ratio into
$\gamma\gamma$ relative to the SM plotted against the inclusive cross
section normalised to the SM for either $h=H_1$ or $h=H_2$. As the
inclusive production is dominated by gluon 
fusion, we can restrict our discussion to this production process. The
figures show that for vanishing $A_t$ gluon fusion can indeed be
enhanced compared to the SM due to stop loop contributions, as has
been discussed in 
Section~\ref{sec:gluonfusion}. With rising mixing the stop loop
contribution interferes destructively, and for $A_t=1$~TeV the gluon
fusion process is suppressed compared to the SM. Also the branching
ratio into photons shows the expected opposite behaviour. For large
values of $A_t$, where constructively interfering stop loops enhance
the partial width, we can observe slightly larger branching ratios than for
$A_t=0$~GeV. It should be kept in mind though that the behaviour of
the branching ratio is an interplay of the partial width into photons
and the total width. Once again for $\tan\beta=4$ the corresponding
plots to Fig.~\ref{fig:gamgamrate} show a similar behaviour with
altogether less parameters points. \sn

Above the black line the reduced cross section $R_{\gamma\gamma}
= R^{BR}_{\gamma\gamma} \, R_{\sigma_{incl}} \ge 1$.\footnote{Note
  that we discuss here the reduced cross section for $h$ only. Later
  we will look at reduced cross sections $\mu_{XX}$ in 
the final state $X$, built up by the 126~GeV Higgs boson and possibly
nearby Higgs resonances. This is what actually is observed in the experiment.}
As can be inferred from the plots, in the NMSSM both $H_1$ and $H_2$ are compatible with
a 126~GeV Higgs boson and an enhanced rate into photon final
states. For $h=H_2$ there are substantially more (red) points, which are
compatible with the constraints that come from the requirement of the
validity of the perturbation theory up to the GUT scale, than for $h=H_1$ (green
points). In particular for vanishing $A_t$ extra matter is required, behaviour
which can be traced back to the need of the 
$H_1$ tree-level mass being as large as possible, {\it cf.}~the
discussion in the previous subsection. We note that there are
scenarios where both the branching ratio and the inclusive production
are very small due to $h$ being very singlet-like. These scenarios
passed the constraint (\ref{eq:cond}) as in this case the photon
reduced cross section $\mu_{\gamma\gamma}$, which can be a
superposition of contributions from various Higgs 
bosons being close in mass, is dominated by the contribution from
another light Higgs boson with a mass of $\sim 126$~GeV, which is not
singlet-like in this case.  \sn

\begin{figure}[t]
 \includegraphics[width=8cm]{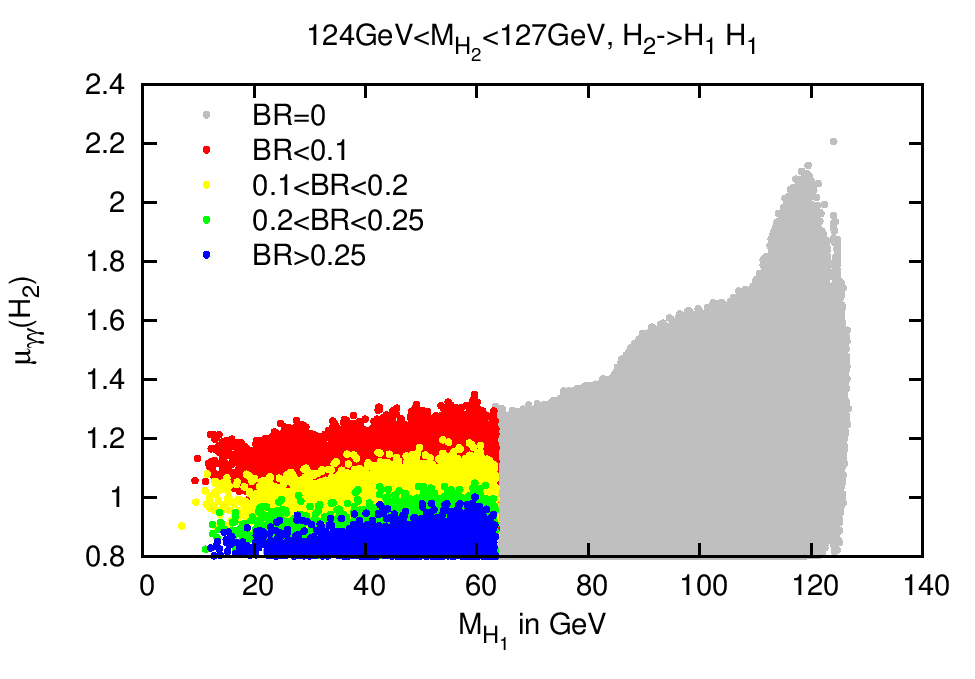}\includegraphics[width=8cm]{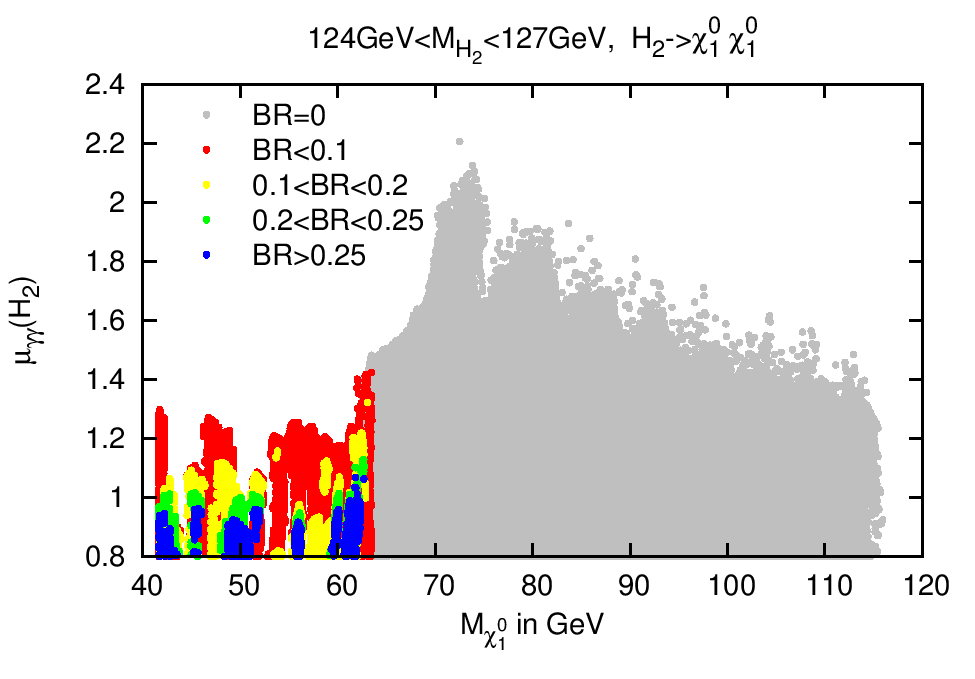}
\caption{Reduced cross sections into $\gamma\gamma$ against $M_{H_1}$
  (left) and $M_{\tilde{\chi}^0_1}$ (right) for $h=H_2$, $\tan\beta=2$ and
  $A_t=1$~TeV. The colour code denotes the size of the branching
  ratio $BR (H_2 \to H_1 H_1)$ (left) and $BR(H_2\to \tilde{\chi}_1^0
  \tilde{\chi}_1^0)$ (right).}
\label{fig:h2decays}
\end{figure}
As already mentioned above the heavier CP-even Higgs boson $H_2$ can
decay into a pair of lighter Higgs bosons or neutralinos in certain parameter
regions. This is shown in Fig.~\ref{fig:h2decays}, where for
$\tan\beta=2$ and $A_t=1$~TeV the
reduced cross section in the $\gamma\gamma$ final state in case of
$h=H_2$, $\mu_{\gamma\gamma} (H_2)$, is plotted against the mass of
the lightest scalar Higgs boson $H_1$ and the mass of the lightest neutralino
$\tilde{\chi}_1^0$, respectively. The colour code denotes the size of
the respective branching ratio, which is zero above the kinematic
thresholds. These rainbow plots show that in case of enhanced photonic
rates such non-standard Higgs decays always remain below about 10--20\%.
The reduced cross section $\mu_{\gamma\gamma}$ is suppressed in case
of sizeable branching ratios above $\sim 0.25$ with a maximum of 
$BR^{\scriptsize \mbox{max}}_{H_2} (H_1H_1) \approx 0.36$ and
$BR^{\scriptsize \mbox{max}}_{H_2} (\tilde{\chi}^0_1 \tilde{\chi}^0_1)
\approx 0.43$. They are small enough not to be excluded by the present experimental
bounds. As can be read off Fig.~\ref{fig:h2decays} (left), 
the largest enhancements in the photon final state occur for almost
degenerate $H_1$ and $H_2$ masses, which corresponds to neutralino
masses around 73~GeV, see Fig.~\ref{fig:h2decays} (right). We
explicitly verified that here the enhanced rate in the photon
final state is due to the increased branching ratio into photons
because of suppressed $H_2$ couplings to $b$ quarks in this case. Due
to sum rules the $H_1$ coupling to $b$ quarks is then substantial. The
combination of the effects of Higgs couplings to SM particles and experimental
exclusion limits then implies the observed pattern in the plots.
Concerning $H_1$, it mainly decays into $b$-quark pairs with a branching ratio of
0.8--0.9, followed by decays into $\tau$ pairs and a branching ratio
of roughly 0.1. The Higgs-to-Higgs or Higgs-to-neutralino decays hence
lead to interesting final
state signatures with {\it e.g.} $4b$, $2b \, 2\tau$, $4\tau$ or even multi-$\mu$ final
states in the former case, from the secondary Higgs decays. In the
latter case the final state lightest neutralino entails large missing
energy. Such events could act as smoking gun signatures for extended
Higgs sectors beyond the minimal SUSY version. 

\subsection{Compatibility with the LHC Higgs search results}
\begin{figure}[t]
\vspace*{-1cm}
 \includegraphics[width=8cm]{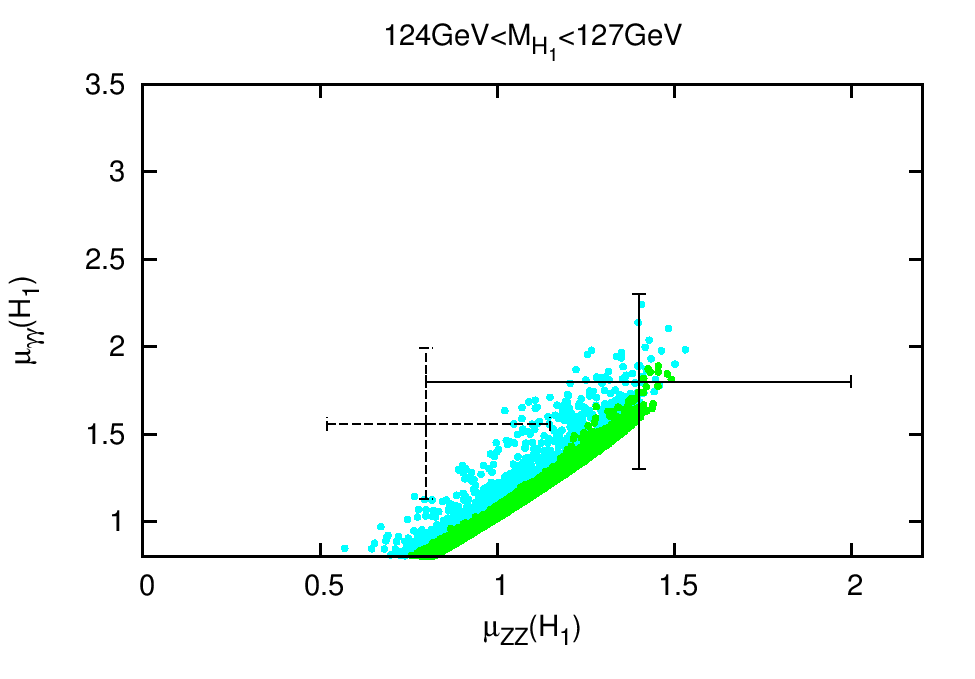}\includegraphics[width=8cm]{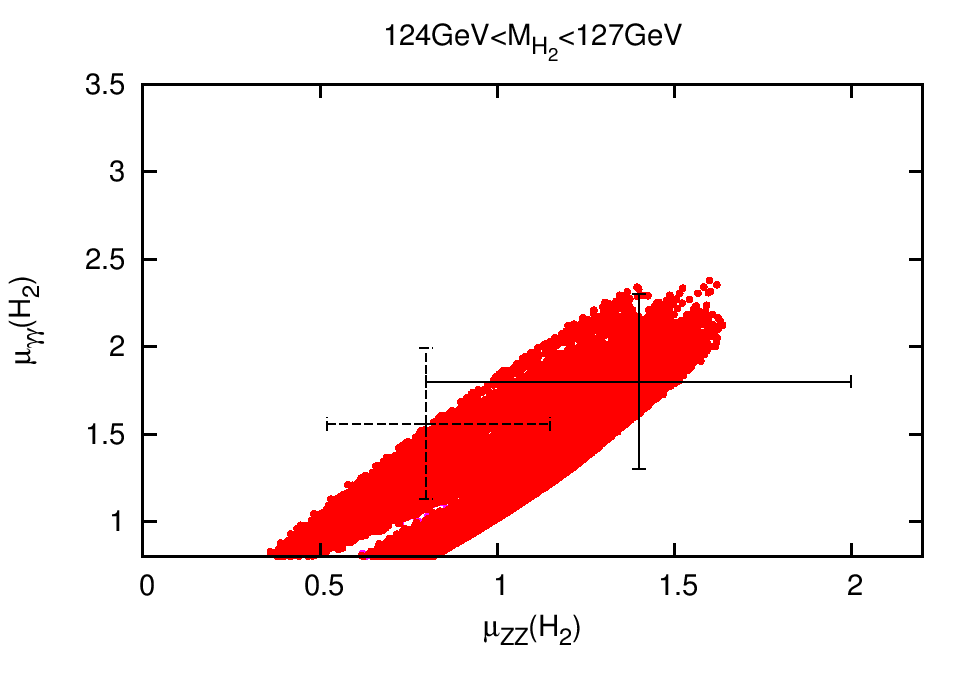}
\includegraphics[width=8cm]{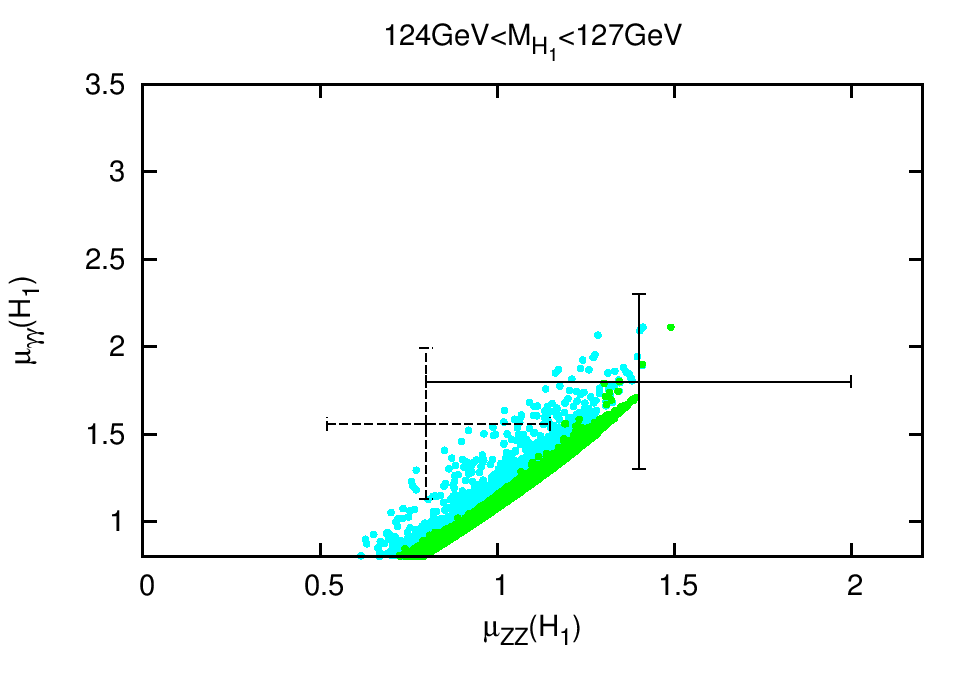}\includegraphics[width=8cm]{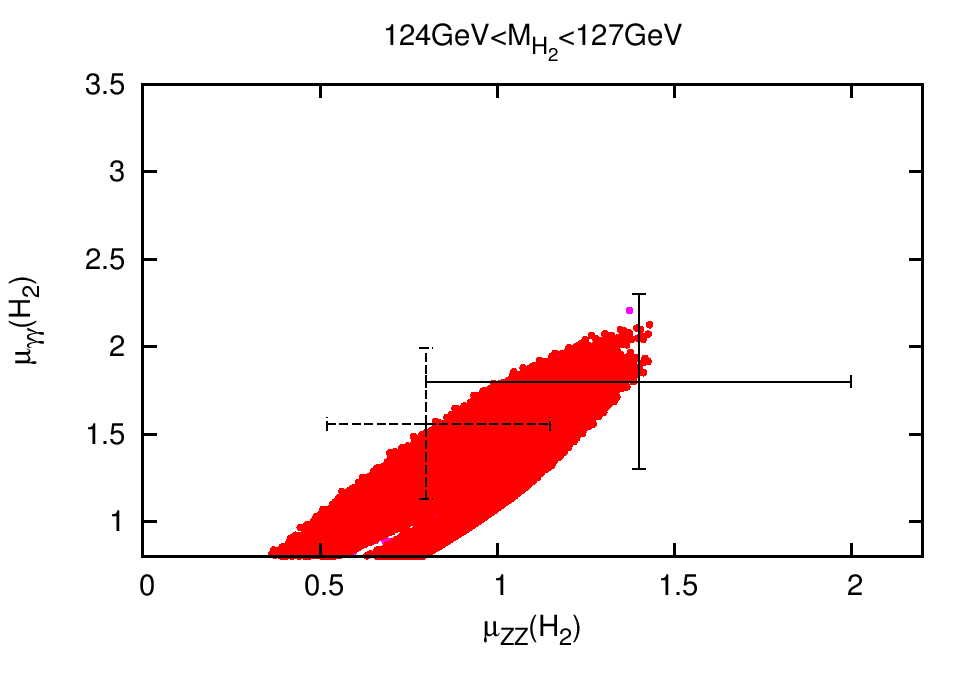}
\caption{Reduced cross section into $\gamma\gamma$ versus the reduced
  cross section into $ZZ$ for $A_t=0$~GeV (upper)
  and $A_t=1$~TeV (lower ) with $h=H_1$ (left) and
  $h=H_2$ (right). Cyan/pink points indicate the signals where at
  least two Higgs bosons with similar masses overlap and the combined
  reduced cross section deviates by more than 10\% from the reduced
  cross section of the individual Higgs boson. Bars: Experimentally
  measured values with error bars (full/ATLAS, dashed/CMS).}
\label{fig:redgauge1}
\end{figure}
\begin{figure}[t]
\vspace*{-1cm}
\includegraphics[width=8cm]{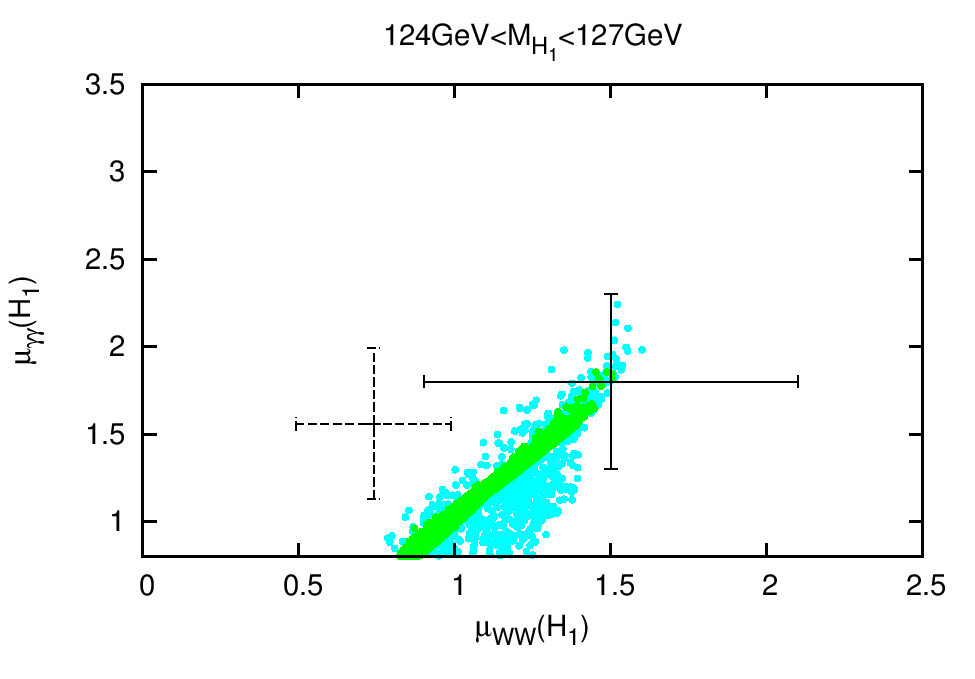}\includegraphics[width=8cm]{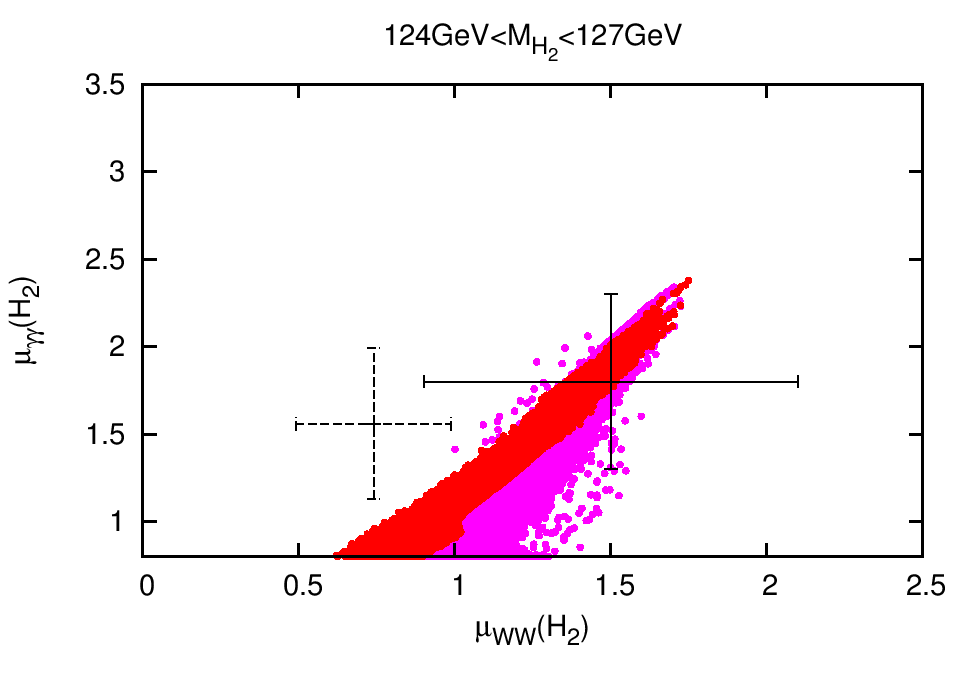}\\[0.2cm]
\includegraphics[width=8cm]{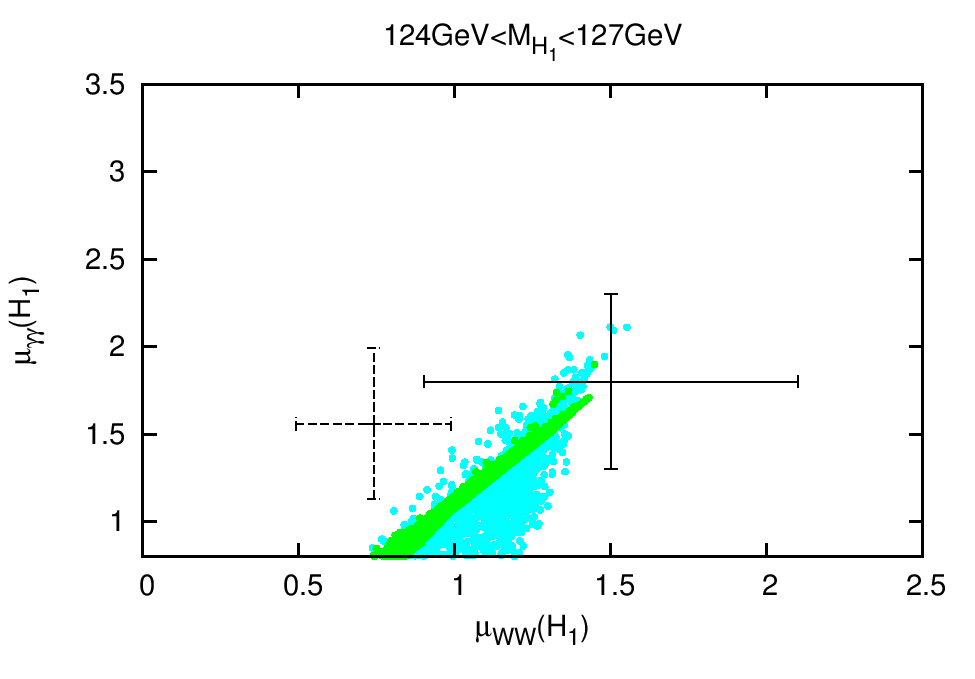}\includegraphics[width=8cm]{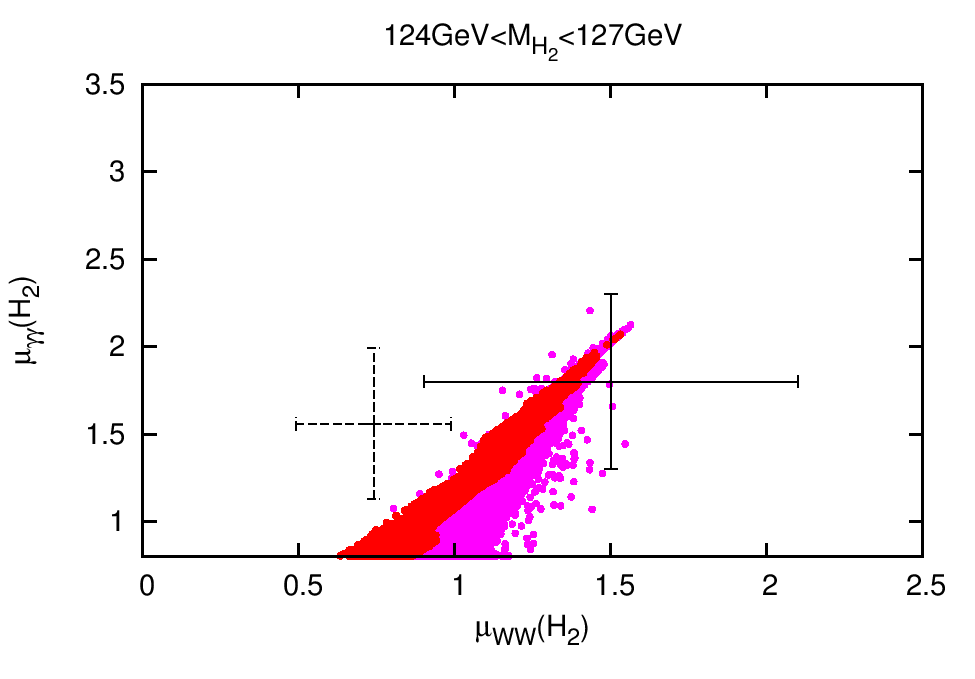}
\caption{Reduced cross section into $\gamma\gamma$ versus the reduced
  cross section into $WW$ for $A_t=0$~GeV (upper)
  and $A_t=1$~TeV (lower) with $h=H_1$ (left) and
  $h=H_2$ (right). Cyan/pink points indicate the signals where at
  least two Higgs bosons with similar masses overlap and the combined
  reduced cross section deviates by more than 10\% from the reduced
  cross section of the individual Higgs boson. Bars: Experimentally
  measured values with error bars (full/ATLAS, dashed/CMS).}
\label{fig:redgauge2}
\end{figure}
\begin{figure}[t]
 \includegraphics[width=8cm]{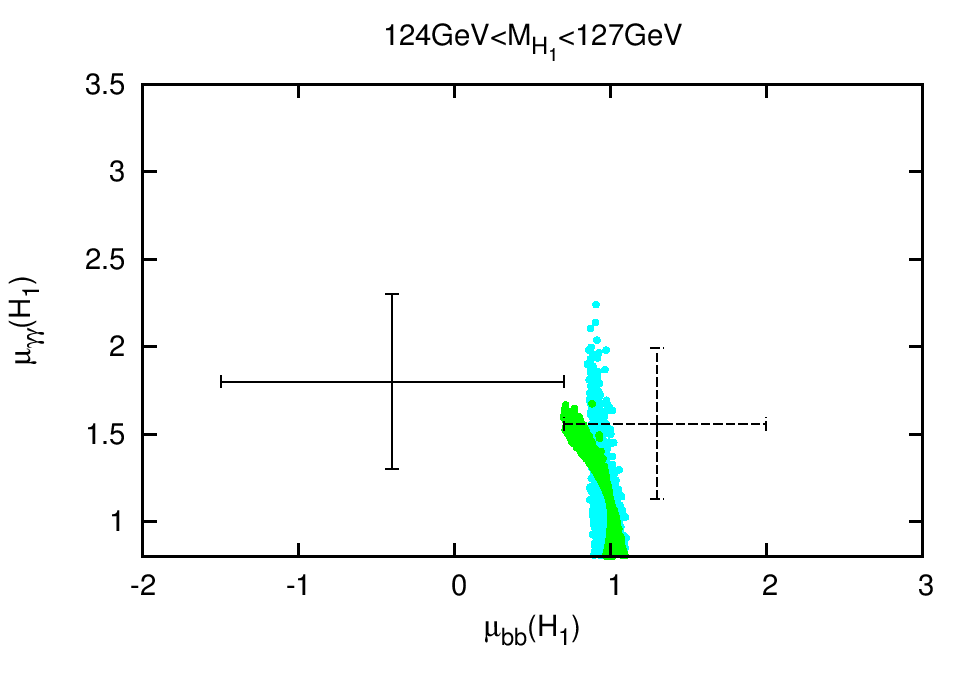}\includegraphics[width=8cm]{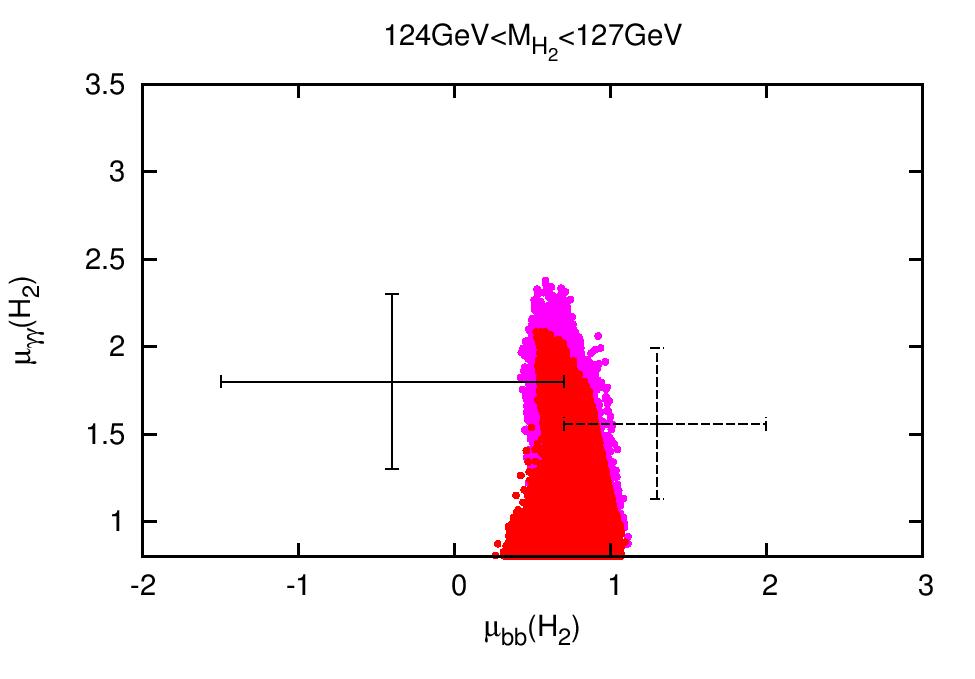}
 \includegraphics[width=8cm]{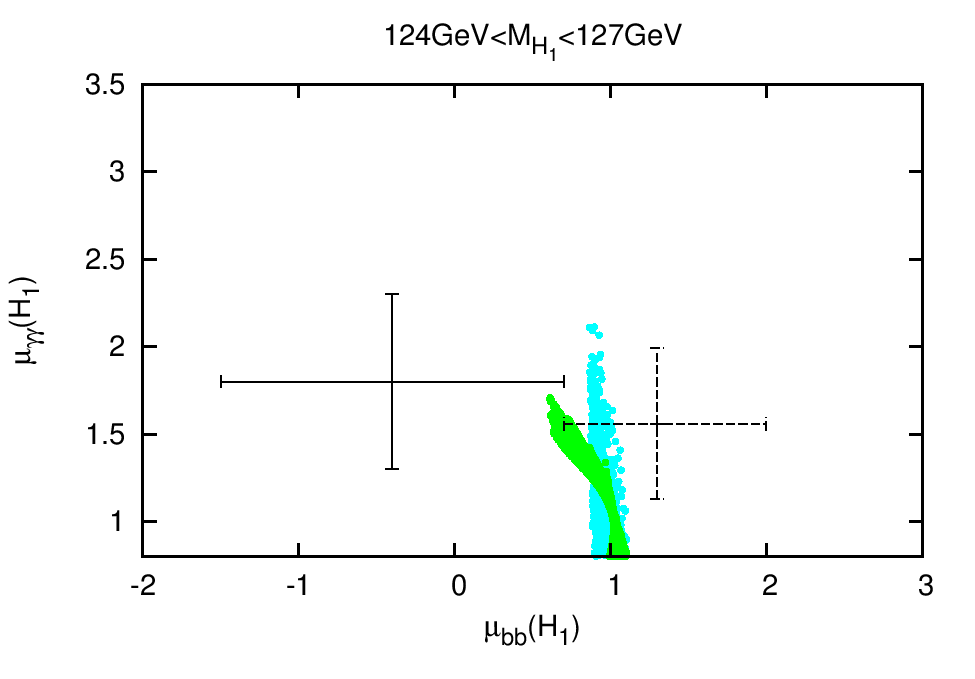}\includegraphics[width=8cm]{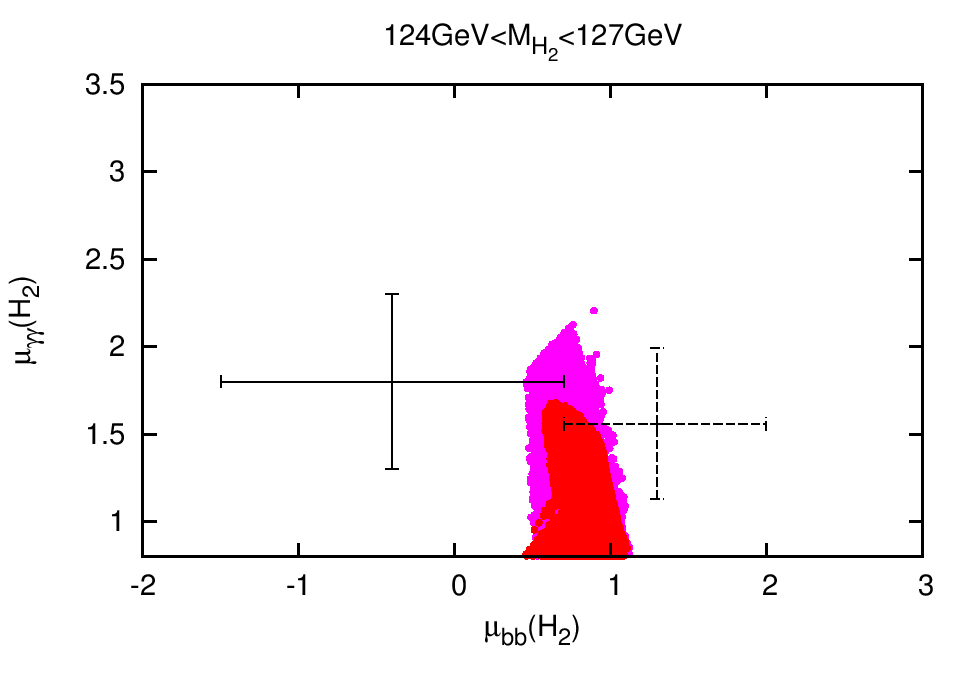}
\caption{Reduced cross section into $\gamma\gamma$ versus the reduced
  cross section into $bb$ for $A_t=0$~GeV (upper)
  and $A_t=1$~TeV (lower) with $h=H_1$ (left) and
  $h=H_2$ (right). Cyan/pink points indicate the signals where at
  least two Higgs bosons with similar masses overlap and the combined
  reduced cross section deviates by more than 10\% from the reduced
  cross section of the individual Higgs boson. Bars: Experimentally
  measured values with error bars (full/ATLAS, dashed/CMS).}
\label{fig:redfermion1}
\end{figure}
\begin{figure}[t]
\includegraphics[width=8cm]{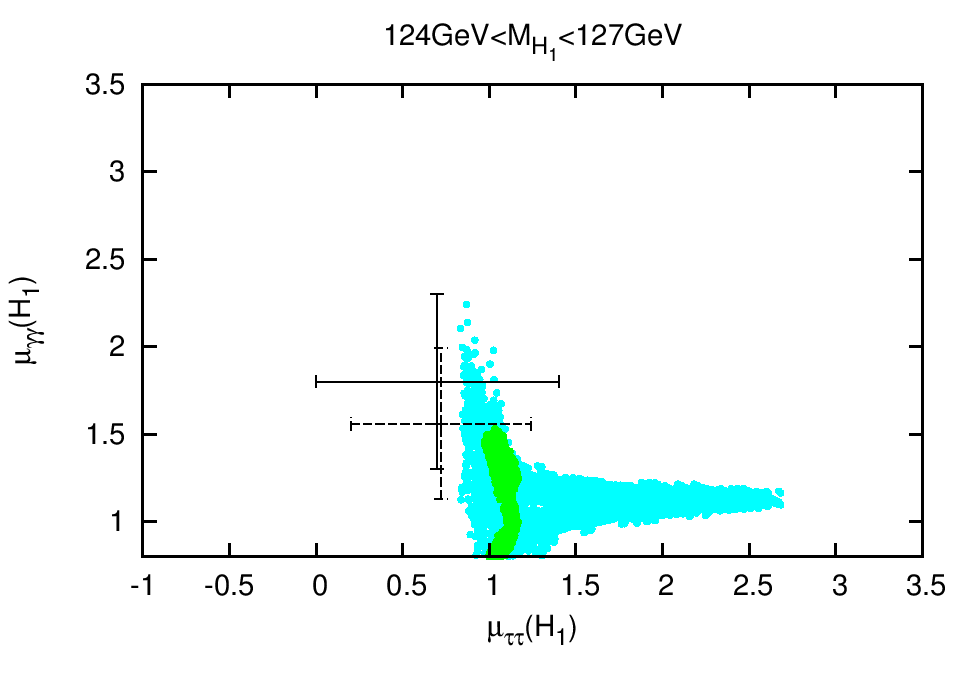}\includegraphics[width=8cm]{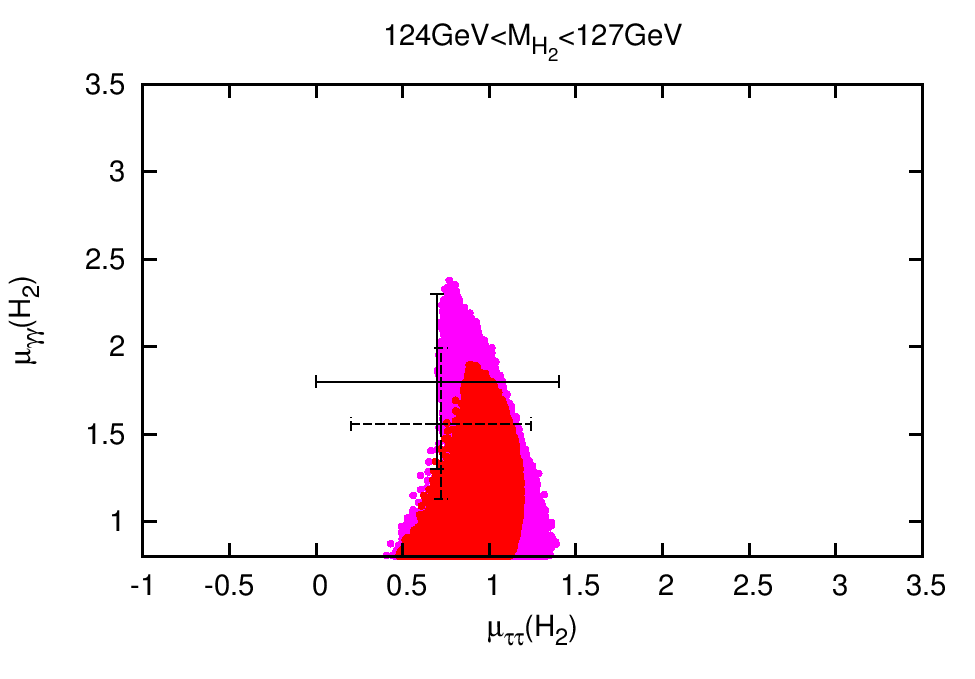} \\[0.2cm]
\includegraphics[width=8cm]{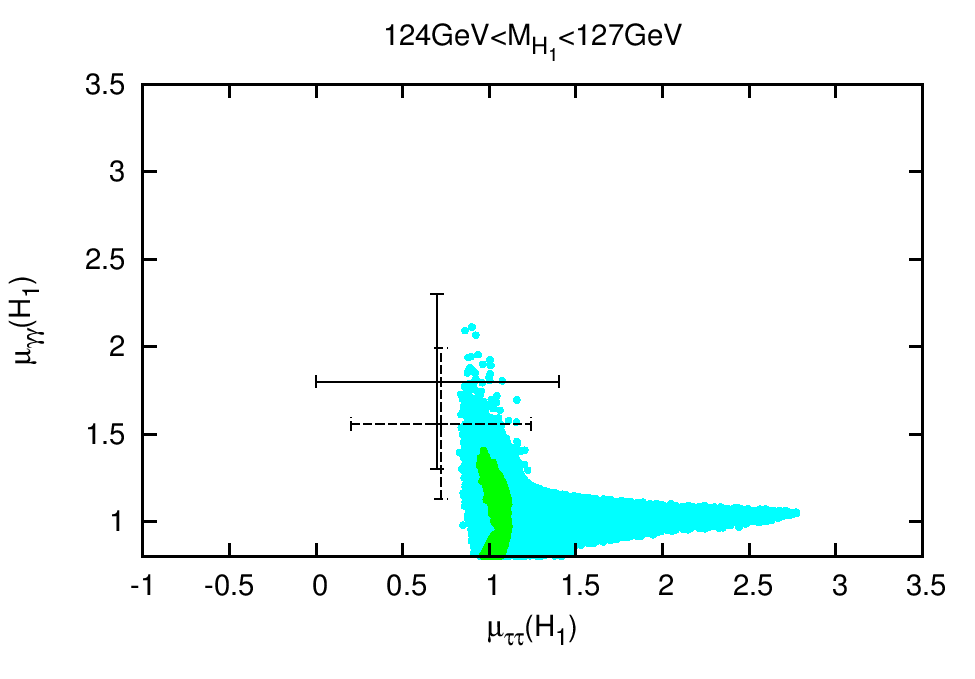}\includegraphics[width=8cm]{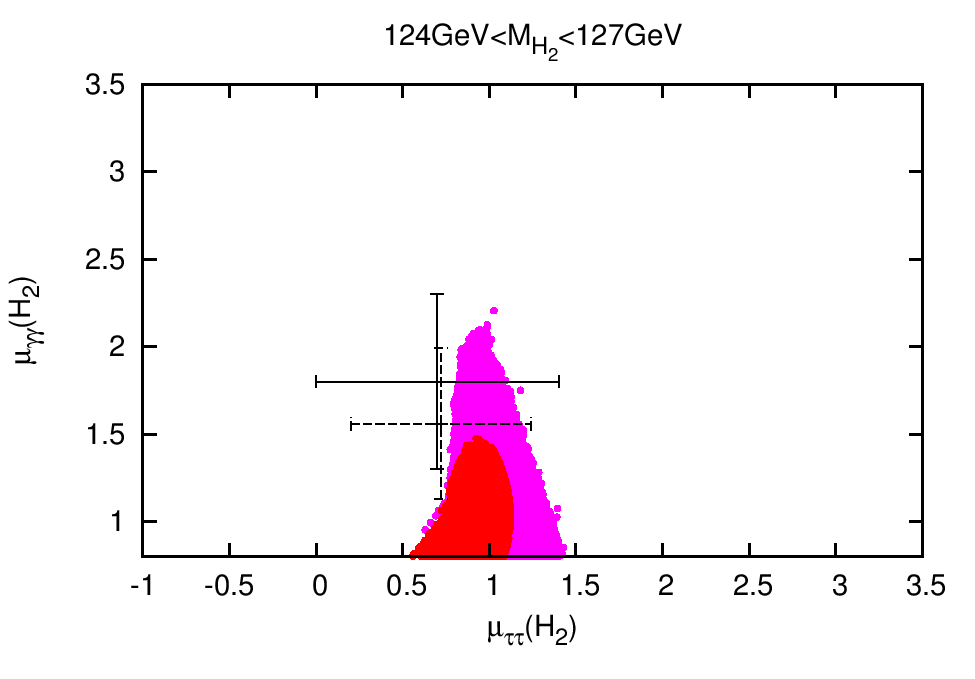} 
\caption{Reduced cross section into $\gamma\gamma$ versus the reduced
  cross section into $\tau\tau$ for $A_t=0$~GeV (upper)
  and $A_t=1$~TeV (lower) with $h=H_1$ (left) and
  $h=H_2$ (right). Cyan/pink points indicate the signals where at
  least two Higgs bosons with similar masses overlap and the combined
  reduced cross section deviates by more than 10\% from the reduced
  cross section of the individual Higgs boson. Bars: Experimentally
  measured values with error bars (full/ATLAS, dashed/CMS).}
\label{fig:redfermion2}
\end{figure}

In this subsection we investigate the compatibility of the results for
the reduced cross sections $\mu_{XX}$ with the experimental best fit
values of the signal strengths in
the various final states. Figures~\ref{fig:redgauge1}--\ref{fig:redfermion2}
show the reduced
cross section in the $\gamma\gamma$ final state compared to the one
in $ZZ$, $WW$, $bb$ and $\tau\tau$, respectively, for $h=H_1$ and $H_2$ with
$A_t=0$~GeV and $A_t=1$~TeV. The bars represent the newest results for
the best fit values of the signal strengths $\mu = \sigma /
\sigma_{\scriptsize \mbox{SM}}$ in the different final states, reported by the ATLAS
\cite{:2012gk,awwkyoto,abbkyoto,atautaukyoto} and the CMS
Collaboration \cite{:2012gu,cwwkyoto,czzkyoto,cbbkyoto,ctautaukyoto},
together with their corresponding errors. The values and errors are
listed in Table~\ref{table:muval} in Appendix~\ref{sec:muval}. First of all the plots
demonstrate that both $H_1$ and $H_2$ can have a mass around 126~GeV and
be compatible with the experiment, for small and for large mixing in the stop
sector. Moreover an enhancement in the photon rate by up to a factor
$\sim 2.4$ is possible. 
The allowed parameter regions are somewhat more
extended for $A_t=0$~GeV, which is an interplay between the production
cross section and decay into photon pairs leading to more important
reduced rates for the small mixing case.
The regions in cyan (pink) indicate where additional Higgs bosons
close in mass join $h=H_1$ ($h=H_2$) to build up the signal and lead
to reduced cross sections that 
differ by more than 10\% from the one of $h$ alone.  
Depending on the value of $A_t$ and the final state these regions are 
more or less extended: The experimental resolution in the various
final states is not the same, which has been taken into account
by applying a different width in the Gaussian smearing of the non $h$ Higgs cross
sections, that are added to the $h$ final state. Therefore the
parameter regions with several Higgs bosons contributing to the final
state are for $WW$ final states, where the Higgs mass cannot
be reconstructed, different from the ones for $ZZ$. The same holds for
the fermionic final states. Here the resolution in the $\tau\tau$
final states is less good than the one in $bb$, leading to the 'nose'
in the plots for $h=H_1$ Fig.~\ref{fig:redfermion2} (left)
against the $\tau\tau$ final state.\footnote{The
  difference in the $bb$ and $\tau\tau$ branching ratios
  due to QCD corrections is small enough not to play a significant
  role here; nor do the negligible $\Delta_b$ corrections.} 
Another reason for the difference in the extensions of the parameter regions is that 
due to the different Higgs-gauge and Higgs-fermion coupling
structures, for a lot of parameter points the non-$h$ Higgs state
contributions to the gauge boson final states can not be important
enough to induce a change in the rate by more than 10\%. 
This is because the Higgs-gauge couplings for small values of $\tan\beta$ are
dominated by the up-type Higgs component. In order to achieve a large
enough production for the $h$ Higgs boson through gluon fusion its
up-type component must be near the SM value, inducing a very small
up-type component for the other CP-even Higgs bosons due to coupling
sum rules, so that they hardly decay into massive gauge bosons. The
down-type component of the Higgs bosons, however, has not been
restricted and therefore both the $h$ Higgs boson and the other
one(s) with mass close by can have equally important couplings to
down-type quarks depending on the amount of singlet-doublet mixing. \sn

A substantial amount of scenarios compatible with an excess in the
photon final state is hence only due to a superposition of Higgs
rates stemming from nearly degenerate Higgs bosons. The experimental
distinction of such scenarios 
from single Higgs rates, as has been discussed {\it e.g.} in
Ref.~\cite{Gunion:2012gc}, would be a clear signal of beyond the SM
Higgs physics. \sn

The plots show the strong correlation between the $\gamma\gamma$ and the massive
gauge boson final states: In case the increase in the photonic final
state is due to an enhanced photon branching ratio caused by a
suppression in the decay width into $bb$, 
this affects the branching ratio into gauge bosons as well and leads also here to
larger rates. Should the gauge boson reduced cross sections turn out to be exactly
SM-like, a strongly enhanced rate into the $\gamma\gamma$ final state
would be difficult to comply with. Nevertheless, even in this case
enhanced photonic rates up to $\sim$1.6--1.8 are still possible. At the present status of
experimental errors and experimental resolution everything
is still compatible. There is a little bit more tension with the CMS
results, as CMS finds suppressed rates  
into $ZZ,WW$ contrary to ATLAS reporting enhanced rates. With more 
data accumulated by the experiments and reduced errors on the $\mu_{XX}$
values future will show which of these scenarios will survive and which
will be excluded. The correlation between the photon and the fermion final
states on the other hand is much less pronounced. While in the gauge
boson final states the branching ratios are simultaneously affected by
a change in the $bb$ decay mode, the down-type fermion final states
are less sensitive to such a change. In the $bb$ final state the $\mu$
value reported by ATLAS lies below the allowed regions, the one of CMS
above, both still compatible within the large errors with the
results of the parameter scan so that at
present no conclusive statement can be made. In the $\tau\tau$ final
state the reported $\mu$ value is below one and hence the
Higgs-$\tau\tau$ coupling suppressed. The ATLAS and CMS values are on
the left border of the allowed parameter range and compatible within
errors, which also in this channel are still too large to make firm statements.
\sn

Concerning perturbativity constraints the regions shown in
Figures~\ref{fig:redgauge1}--\ref{fig:redfermion2} are mostly
compatible within the NMSSM without the inclusion of extra matter. A
few scenarios require the inclusion of extra matter at 1~TeV. 
In case of $\tan\beta=4$ the shapes of the parameter regions
corresponding to Figures~\ref{fig:redgauge1}--\ref{fig:redfermion2}
stay approximately the same but are much less dense in the amount of
allowed scenarios. \sn 

With increasing data the precision on the signal strengths reported by
the experiments will improve and the exclusion limits will become more
stringent. This has to be taken into account when combining
signals stemming from two Higgs bosons which are close in mass. Thereby the
reduced cross sections $\mu_{XX}$ will change. In particular the enhancement
in the $\gamma\gamma$ final state, which in a substantial amount of
scenarios is due to this superposition, may partially disappear. Furthermore, it affects
scenarios which have been excluded due to too large signal rates from the combination
of two Higgs signals. On the other hand, the improved precision would then
reveal two Higgs signals lying close to each other, providing an unambiguous sign
of BSM physics, should the NMSSM or some other multi-Higgs sector be realised in nature.

\section{Summary and Outlook \label{sec:concl}}
In this paper we have studied the phenomenology of Higgs bosons close to 126~GeV
within the scale invariant unconstrained NMSSM, focusing on the
regions of parameter space favoured by low fine-tuning considerations,
namely stop masses of order 400~GeV to 1~TeV and an effective $\mu$
parameter between 100--200~GeV, with large $\lambda$ (which is
required to remain perturbative up to the GUT scale) and low $\tan
\beta =$2--4.  \sn

By performing scans over the above parameter space, focusing on the
observable Higgs cross sections into $\gamma \gamma$, $WW$,  $ZZ$,
$bb$ and $\tau \tau$ final states, we have studied the correlations between these observables.
Although we examined only a limited parameter range in $A_\kappa$, $A_\lambda$ and
$\kappa$, we found a substantial amount of parameter space which can lead
to Higgs boson masses and couplings compatible with the latest LHC results. \sn

There are basically two types of NMSSM scenarios compatible with the data,
corresponding to the 126~GeV Higgs boson being either the lightest
CP-even Higgs boson $H_1$ or the second lightest one $H_2$.
Our results clearly favour the second option, however the first option is 
still possible but it requires additional extra matter at the TeV scale in order to 
maintain the perturbativity of $\lambda$, as well as  large stop
mixing and low $\tan\beta \sim 2$  (for example $\tan\beta=4$ is not
allowed) and even then the allowed parameter space is relatively
sparse. We emphasise that these conclusions only apply to the natural
NMSSM, in the low fine-tuning region defined above, and that larger
stop masses and mixing (above one TeV) would allow a larger parameter
space with $H_1$ at 126~GeV. \sn

The enhancement in the Higgs rate in the di-photon channel that we
observe in our results is due to a combination of factors. Firstly
there can be an enhancement in the dominant gluon fusion Higgs
production cross section due to the light squarks in the loop,
where light stops are a feature of the low fine-tuning region. 
Secondly the Higgs branching ratio in the di-photon channel can be
enhanced due to two sub-factors, namely (i) an increased di-photon
partial width, induced by stop, charged Higgs and chargino
loops, and (ii) a suppressed total width, due to a suppressed Higgs
coupling to $b$ quarks resulting from singlet-doublet
mixing. Concerning case~(i), the stop loop effect in the photonic
decay is opposite to the one in gluon fusion and depends on the mixing.  In the case
(ii) also the rates into the fermionic final states can be
suppressed. The reported best fit values of the signal strengths $\mu$
in the $bb$ and $\tau\tau$ final states by ATLAS and CMS still suffer
from large errors, so that it is difficult to draw a conclusion 
on possibly suppressed couplings to down-type fermions. The allowed parameter ranges we
found for enhanced di-photon event rates are compatible with the
experimental best fits of the $\mu$ values in the various final states. \sn

Although our results encompass the SM case where all $\mu$ values are
equal to unity, we also allow for significant and correlated departures from unity for all channels.
While we do not find any significant correlation between di-photon and
fermion rates ($\mu$ values), we do find a correlation between the
di-photon and massive gauge boson rates  ($\mu$ values). 
However, given the present status of the experimental accuracy in the various final
states, it is not possible to draw any conclusions about this.
Nevertheless it is clear that the natural NMSSM Higgs sector
(corresponding to the low fine-tuning region as 
defined above) is nicely compatible with all experimental results,
with the bulk of the data points corresponding to the  
the second lightest Higgs boson $H_2$ having a mass of about 126~GeV.
In this favoured case, the 126~GeV $H_2$ boson can decay into pairs of
lighter neutralinos, CP-even or CP-odd Higgs bosons, providing a
smoking gun signature of the NMSSM. \sn 

We also emphasise that a good part of the parameter space involves 
a Higgs spectrum where the two lighter CP-even Higgs bosons $H_1$ and $H_2$ are
close in mass. It is then the combination of their reduced cross sections and rates which
is observed in the experiment. However, with increasing accuracy in
the Higgs boson mass resolution, future LHC data may resolve these two states.
Observing the two separate CP-even Higgs bosons $H_1$ and $H_2$ with different masses
would not only rule out the Standard Model, but could also provide direct evidence for the 
Higgs sector of the NMSSM. \sn

With future LHC results, the best fit values of the signal strengths
and their errors in the different channels will
change, leading to different positions and error bars on the data
points represented by crosses in our plots. However, the overall
pattern of the plots themselves will not change substantially.
Thus future data can be compared to our predictions to check the
compatibility of the natural NMSSM with experiment. If stops are not
discovered below one TeV, and instead the experimental limit on the stop
masses increases, then the range of the stop masses and mixing may
need to be extended beyond the low fine-tuned region considered here,
leading to enlarged parameter regions of the NMSSM. However, for the
moment, the natural NMSSM is still viable, with the characteristic Higgs spectrum and
properties discussed in this paper.

\subsubsection*{Acknowledgments} 
MMM and KW are supported by the DFG SFB/TR9  ``Computational Particle Physics''. 
SFK is supported by the STFC Consolidated grant ST/J000396/1 and EU ITN grants UNILHC 237920 and INVISIBLES 289442 . We would like to thank G\"unter Quast and Markus
Schumacher for very helpful discussions. MMM furthermore thanks Wim de Boer for
interesting conversations on NMSSM and Dark Matter. RN is grateful to
M.I. Vysotsky and L.B. Okun for fruitful discussions. \newpage

\section*{Appendix}
\begin{appendix}

\section{\label{sec:muval} Best fit values of the signal strength}
We list the best fit values of the signal strengths
$\mu=\sigma/\sigma_{\scriptsize \mbox{SM}}$ in the various
final states reported by ATLAS
\cite{:2012gk,awwkyoto,abbkyoto,atautaukyoto} and CMS
\cite{:2012gu,cwwkyoto,czzkyoto,cbbkyoto,ctautaukyoto}, which we have
applied in our plots.\footnote{The statistical ($\pm 0.7$) and
  systematic error ($\pm 0.8$) in the $bb$ final state reported by
  ATLAS have been added in quadrature.} 

\renewcommand{\arraystretch}{2}
\begin{table}[!h]
  \centering
  \begin{tabular}{|c|l|c|c|}
    \hline
 Experiment & Final state & $(\sqrt{s},\mbox{ L})$  & $\mu =
 \sigma/\sigma_{\scriptsize \mbox{SM}}$ \\
 \hline \hline
ATLAS & $\gamma\gamma$ &  (7~TeV, 4.8~fb$^{-1}$)+(8~TeV,
5.9~fb$^{-1}$) & $1.8 \pm 0.5$ \cite{:2012gk} \\
 & $WW$ & (8~TeV, 13~fb$^{-1}$) & $1.5\pm 0.6$
 \cite{awwkyoto} \\
 & $ZZ$ & (7~TeV, 4.8~fb$^{-1}$)+(8~TeV, 5.8~fb$^{-1}$) &
 $1.4\pm 0.6$ \cite{:2012gk} \\
 & $bb$ & (7~TeV, 4.7~fb$^{-1}$)+(8~TeV, 13~fb$^{-1}$) & $-0.4\pm
 1.1$\cite{abbkyoto} \\
 & $\tau\tau$ &  (7~TeV, 4.6~fb$^{-1}$)+(8~TeV, 13~fb$^{-1}$) & $0.7
 \pm 0.7$ \cite{atautaukyoto}\\ \hline\hline
CMS & $\gamma\gamma$ &  (7~TeV, 5.1~fb$^{-1}$)+(8~TeV,
5.3~fb$^{-1}$) & $1.56 \pm 0.43$ \cite{:2012gu} \\[0.1cm]
 & $WW$ & (7~TeV, 4.9~fb$^{-1}$)+(8~TeV, 12.1~fb$^{-1}$) & $0.74\pm 0.25$
 \cite{cwwkyoto} \\
 & $ZZ$ & (7~TeV, 5.1~fb$^{-1}$)+(8~TeV, 12.2~fb$^{-1}$) &
 $0.8^{+0.35}_{-0.28}$ \cite{czzkyoto} \\
 & $bb$ & (7~TeV, 5~fb$^{-1}$)+(8~TeV, 12~fb$^{-1}$) & $1.3^{+0.7}_{-0.6}$ \cite{cbbkyoto} \\
 & $\tau\tau$ & (7~TeV+8~TeV, 17~fb$^{-1}$) & $0.72 \pm 0.52$
 \cite{ctautaukyoto} \\ \hline
\end{tabular}
  \caption{Best fit values of the signal strength $\mu$ in the
    $\gamma\gamma$, $WW$, $ZZ$, $bb$ and $\tau\tau$ final states
    reported by ATLAS \cite{:2012gk,awwkyoto,abbkyoto,atautaukyoto}
    and CMS
    \cite{:2012gu,cwwkyoto,czzkyoto,cbbkyoto,ctautaukyoto}.}
  \label{table:muval}
\end{table}

\end{appendix}

 
\end{document}